\documentclass[pra,aps,amssymb,twocolumn,noshowpacs,hyperref]{revtex4-2}
\usepackage{color}
\usepackage{graphicx}
\usepackage{hyperref}
\usepackage{amsmath}
\usepackage{latexsym}
\usepackage{amssymb}
\usepackage{mathrsfs}
\usepackage{mathtools}
\usepackage{layout}
\usepackage{verbatim}
\usepackage{multirow}
\usepackage{bm}
\usepackage{amsfonts,epsfig}

\begin{document}
\title{Coherence-preserving cooling of nuclear spin qubits in a weak magnetic field}

\date{\today}
\author{Xiao-Feng Shi}
\affiliation{School of Physics, Xidian University, Xi'an 710071, China}

\begin{abstract}
Nuclear spin memories of divalent neutral atoms can allow spin-preserving resolved-sideband cooling in a strong magnetic field [Phys. Rev. Lett. 99, 123001 (2007)]. We present a theory for cooling $^{87}$Sr nuclear-spin qubits in a weak magnetic field. The theory depends on laser excitation of $5s5p~^1P_1$ to a nearby state which results in $m_J$-dependent AC Stark shifts large compared to the hyperfine interaction. This effectively suppresses the nuclear-spin mixing due to the hyperfine interaction. Sideband cooling via the clock state quenched by the AC Stark-shifted $^1P_1$ state leads to nuclear-spin-preserving spontaneous emission back to the ground state. More than being compatible with low magnetic fields, the theory is applicable when the nuclear spin qubits are defined by the two lowest Zeeman substates.

\end{abstract}
\maketitle

\section{introduction}\label{sec01}
Long-lived quantum registers provide a favorable setting for large-scale quantum computing~\cite{Nielsen2000}. Physical systems studied for this purpose include superconducting circuits and trapped ions, where the number of qubits can be up to about 50 in one register~\cite{Zhang2017,Arute2019,Neill2021}, and the number of typical gate operations~(such as Bell-state creation) within the qubit lifetime is on the order of $10^3$ and $10^6$ for superconducting circuits and trapped ions, respectively~\cite{Shi2021qst}.

Recently, quantum registers with over 200 neutral-atom qubits~\cite{Schymik2020,Semeghini2021,Ebadi2021,Graham2022} were experimentally realized for coherent quantum control. The long lifetime~\cite{Wang2016,Young2020} of atomic qubits and fast entangling operations~\cite{Madjarov2020} suggest that neutral atoms are leading candidates for quantum memories. For the widely used alkali-metal atoms where qubits are encoded in hyperfine states~\cite{Shi2021qst}, however, heating effects inevitably require recooling of the atoms. Standard laser cooling methods will destroy the quantum information stored~\cite{Saffman2010,Saffman2016} limiting the total number of quantum gates that can be executed within the register lifetime. To prolong the memory lifetime effectively, coherence-preserving cooling of alkali-metal atoms was proposed by resorting to superfluid immersion~\cite{Daley2004},~cavity QED~\cite{Griessner_2004}, or coupling qubits to auxiliary atoms~\cite{Belyansky2019}.

When the qubits are defined by the nuclear spin states of alkaline-earth-like (AEL) atoms, including alkaline-earth metals, some lanthanides~\cite{Robicheaux2017} and some transition metals, resolved-sideband cooling may preserve the nuclear spin coherence in the presence of sufficiently strong magnetic fields~\cite{Reichenbach2007}. With $I$ and $m_I$ the nuclear spin and its projection along the quantization axis, numerical analyses in Ref.~\cite{Reichenbach2007} showed that for qubits defined with $\pm m_I$ in the ground state of $^{87}$Sr and $^{117}$Yb, where $0<m_I\leq I$, spontaneous emission during cooling can preserve the qubit-state coherence with a fidelity over 0.99 in a strong magnetic field. The B-field is about $10$~mT to achieve a fidelity over 0.99 for $^{87}$Sr. 

In this paper we propose resolved-sideband cooling of $^{87}$Sr atoms in a weak magnetic field while preserving the coherence of nuclear spin qubits. Following Ref.~\cite{Robicheaux2017}, we consider a cooling cycle in which the ground state is driven to the vibrational sideband of the clock state, which is further driven to the $(5s5p) ^1P_1$ state, which decays rapidly back to ground. In our theory, the hyperfine-interaction-induced mixing of different nuclear spin states in $|[5s5p ^1P_1]m_J,m_I\rangle$  is effectively suppressed by coupling it to a nearby state which causes $m_J$-dependent AC Stark shifts. In particular, the Stark shift is large compared to the hyperfine interaction, so that the nuclear spin mixing due to the hyperfine interaction becomes negligible. This mechanism is not dependent on the Zeeman shift, which leads to two features. First, a weak magnetic field is applicable, which is compatible with recent nuclear-spin-qubit experiments, where a B-field of $11$~G~\cite{Barnes2022}, $4.11$~G~\cite{Ma2022}, or a value in the range $(0,~18]$~G~\cite{Jenkins2022} was used with $^{87}$Sr~\cite{Barnes2022} or $^{171}$Yb~\cite{Ma2022,Jenkins2022}. Second, the theory is for qubits defined with the two lowest nuclear spin Zeeman substates, which was commonly used in experiments, such as in the experiment of Ref.~\cite{Barnes2022}. Numerical simulations with feasible parameters show that nuclear spin coherence can be preserved with a fidelity over $0.999$. This theory brings opportunities for coherent control of nuclear-spin quantum memories~\cite{Daley2008,Gorshkov2009,Omanakuttan2021,Shi2021-2,Chen2022,Wu2022}.

\begin{figure*}
\includegraphics[width=6.0in]
{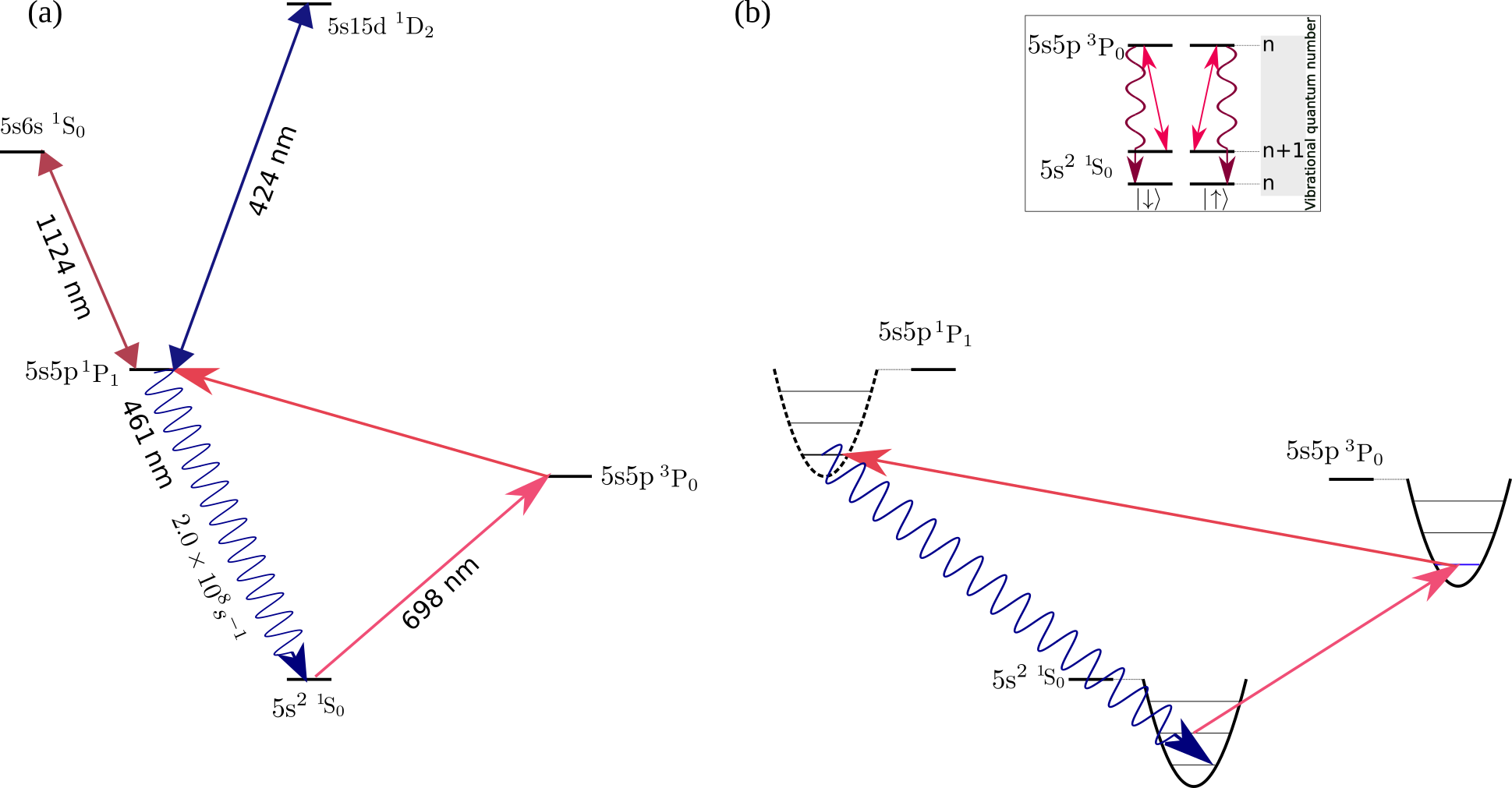}
\caption{Cooling scheme illustrated with atomic levels and vibrational states in (a) and (b), respectively; the inset of (b) shows a simplified process about removing one vibrational quantum number. Altogether five energy levels are involved in the nuclear-spin-preserving sideband cooling of $^{87}$Sr atoms, namely, the ground state, the clock state, $5\text{s}5\text{p}~^{1}\text{P}_{1}$, $5\text{s}6\text{s}~^{1}\text{S}_{0}$, and $5\text{s}15\text{d}~^{1}\text{D}_{2}$. The directions of the arrows do not indicate polarization of photons. Cooling starts from a narrow-line laser excitation of the clock transition from the ground state to $5\text{s}5\text{p}~^{3}\text{P}_{0}$ when the vibrational quantum number reduces by one. A two-photon transition via an intermediate state~(not shown here) transfers the state $5\text{s}5\text{p}~^{3}\text{P}_{0}$ to $5\text{s}5\text{p}~^{1}\text{P}_{1}$ which decays back to the ground state rapidly. The hyperfine interaction in $5\text{s}5\text{p}~^{1}\text{P}_{1}$ mixes nuclear spins by nature; by inducing a transition between $5\text{s}5\text{p}~^{1}\text{P}_{1}$ and $5\text{s}6\text{s}~^{1}\text{S}_{0}$ with a strong Rabi frequency which is large compared to the hyperfine interaction, the nuclear spin mixing is suppressed, so that polarization resolution is removed in the spontaneous emission to the ground state. A small diagonal-hyperfine-interaction induced energy difference between the two nuclear spin states is compensated by the AC Stark shift via off-resonantly exciting $5\text{s}5\text{p}~^{1}\text{P}_{1}$ to $5\text{s}15\text{d}~^{1}\text{D}_{2}$, which removes the frequency resolution in the spontaneous emission.   \label{figure1} }
\end{figure*}

The remainder of this paper is organized as follows. In Sec.~\ref{sec02}, we discuss sideband cooling when ignoring the hyperfine interaction as a warm-up. In Sec.~\ref{sec03}, we present the theory of using AC Stark shifts to suppress the hyperfine interaction. Section~\ref{sec04} shows the detail with a concrete model and presents numerical results of the cooling. Section~\ref{Sec05} discusses the influence from fluctuation of laser frequency, intensity, and polarization on the cooling. Section~\ref{Sec06} gives a discussion especially on the possibility to apply the cooling scheme with other elements, and a brief conclusion is given in Sec.~\ref{Sec07}.

\section{When there is no hyperfine interaction}\label{sec02}
With $^{87}$Sr as an example, the essence of nuclear-spin-preserving resolved sideband cooling in a weak magnetic field is understood by first ignoring the hyperfine interactions in $(5s5p)^1P_1$. The full treatment of hyperfine interaction will be shown in Sec.~\ref{sec03}. 

The cooling consists of three steps. 

First, a narrow-line $698$~nm laser field coherently excites the ground state to the clock state when the vibrational quantum number reduces by one. See Fig.~\ref{figure1}. The electron-nuclear spin state and the vibration state of a ground-state $^{87}$Sr atom is denoted by
\begin{eqnarray}
|[5\text{s}^2~^{1}\text{S}_{0}]m_I\rangle\otimes |n+1\rangle, \nonumber
\end{eqnarray}
where $m_I$ is the nuclear spin projection along the quantization axis~(specified by an external magnetic field $B\mathbf{z}$), and $|n+1\rangle$ denotes the vibrational state of the atom in the trap with $n+1$ vibrational quantum number. We suppose that the clock state and the ground state are simultaneously trapped in a trap of magic wavelength~\cite{Brusch2006,Barnes2022}, so that the vibration states of the atom in the ground and clock states can be denoted by the same set of vibrational states $|n\rangle$. The linewidth for the atomic electric dipole transition from $5\text{s}^2~^{1}\text{S}_{0}$ to $5\text{s}5\text{p}~^{3}\text{P}_{0}$ is about $2\pi\times0.001$~Hz~\cite{Boyd2007}, while the (radial) frequency of the trap can be significantly larger than the transition linewidth; for example, it was $2\pi\times95$~kHz and $2\pi\times260$~kHz in the experiment of Ref.~\cite{Barnes2022} and Ref.~\cite{Brusch2006}, respectively. As a result, the sideband of the vibrational states can be well resolved in the laser excitation of the clock transition
\begin{eqnarray}
|[5\text{s}^2~^{1}\text{S}_{0}]m_I\rangle\otimes |n+1\rangle \rightarrow |[5\text{s}5\text{p}~^{3}\text{P}_{0}]m_F=m_I\rangle\otimes |n\rangle\nonumber\\ \label{step1}
\end{eqnarray}
via a $\pi$ polarized laser field. The atomic state $|[5\text{s}5\text{p}~^{3}\text{P}_{0}]m_F\rangle$ can be written as $|[5\text{s}5\text{p}~^{3}\text{P}_{0}]m_I\rangle$ because the hyperfine and spin-orbit coupling result in a state $|[5\text{s}5\text{p}~^{3}\text{P}_{0}]m_F=m_I\rangle\approx|[5\text{s}5\text{p}~^{3}\text{P}_{0}]m_I\rangle +\eta |\text{hyper-so}\rangle$, where the value of $m_F$ is equal to that of $m_I$, $|\eta|^2\approx 4\times10^{-8}$, and $|\text{hyper-so}\rangle$ is a superposition of $|[5\text{s}5\text{p}~^{3}\text{P}_{1}]m_F\rangle$, $|[5\text{s}5\text{p}~^{3}\text{P}_{2}]m_F\rangle$, and $|[5\text{s}5\text{p}~^{1}\text{P}_{1}]m_F\rangle$~\cite{Boyd2007}. The tiny $|\eta|$ indicates that the hyperfine-interaction-induced nuclear spin decoherence in the transition between the ground state and the clock state can be ignored.

Second, a two-photon Raman transition between $|[5\text{s}5\text{p}~^{3}\text{P}_{0}]m_F=m_I\rangle\otimes |n\rangle$ and $|[5\text{s}5\text{p}~^{1}\text{P}_{1}]m_F-1\rangle\otimes |n\rangle$ via an intermediate state is realized with the total change of angular momentum projection equal to -1. The intermediate state should have both singlet and triplet components for which we have two choices. One choice is a high-lying $5\text{s}n\text{s}$ Rydberg state in which the hyperfine interaction can induce mixing between the $^3S_1$ and $^1S_0$ states~\cite{Ding2018,Shi2021}. The other choice is a low-lying  $5\text{s}n\text{d}$ Rydberg state in which the spin-orbit coupling can induce strong mixing between the $^1D_2$ and $^3D_2$ states. For the first choice, the hyperfine interaction in the intermediate state can lead to different Rabi frequencies for the two nuclear-spin states, and it demands efforts to tune the laser frequencies and detunings for addressing the hyperfine-split intermediate state or use multiple laser fields to achieve equal excitation Rabi frequencies for the two nuclear-spin states. For the second choice, sizable singlet-triplet mixing can occur for $5\text{s}n\text{d}$ states of principal quantum number from $n=10$ to $25$~\cite{PhysRevA.47.4725}. For example, the components of $^1D_2$ and $^3D_2$ in the $^3D_2$-dominated wavefunction for $n=11$ have a ratio of about $5.5$~(see Fig.~3 of Ref.~\cite{PhysRevA.47.4725}). The transition from pure $^3P_0$ to $^1D_2$ or $^3D_2$ states is difficult, but the wavefunction of the clock state has an overlap coefficient $-2\times10^{-4}$ with the pure $^3P_1$ state~\cite{Boyd2007}, which makes it possible to couple the clock state to the intermediate $5\text{s}n\text{d}$ state. The hyperfine splittings in the $5\text{s}n\text{d}$ state with $n=11$ are small~(see Figs.~6 and 7 of Ref.~\cite{PhysRevA.47.4725}), so that when we use a Raman transition with the detuning at the intermediate $5\text{s}n\text{d}$ state of the two-photon transition large compared to the hyperfine interaction, the hyperfine structure of the intermediate state is indiscernible. This means that by the second choice with $n\lesssim11$, the Raman transitions for the two nuclear-spin qubit states can have the same Rabi frequencies as required by our theory. The state $5\text{s}5\text{p}~^{1}\text{P}_{1}$ has a linewidth $2\pi\times32$~MHz~\cite{Xu2003,Millen2010}, and is not trapped by the optical trap. When there is no hyperfine interaction for this state~(the full theory with hyperfine interaction is in Sec.~\ref{sec03}), a two-photon $\sigma^-$ transition via an intermediate state can lead to 
\begin{eqnarray} 
&&|[5\text{s}5\text{p}~^{3}\text{P}_{0}]m_F=m_I\rangle\otimes |n\rangle  \nonumber\\ &&\rightarrow
|[5\text{s}5\text{p}~^{1}\text{P}_{1}]m_J=-1,m_I\rangle\otimes |n\rangle,\label{step2}
\end{eqnarray}
where we preserve the vibrational state when the atom is in $5\text{s}5\text{p}~^{1}\text{P}_{1}$ following Ref.~\cite{Reichenbach2007}. To understand this, we note that the time for the atom to stay in $5\text{s}5\text{p}~^{1}\text{P}_{1}$ is about $5$~ns, while the vibration period is over 10~$\mu$s for a radial trap frequency $2\pi\times95$~kHz~\cite{Barnes2022}. Note that the effective motional temperature of the atom was below $5~\mu$K in recent experiments with ytterbium~\cite{Jenkins2022} or strontium~\cite{Norcia2018,Cooper2018,Covey2019}, and we can assume that the atomic temperature is on the order of $10~\mu$K at the beginning of the sideband cooling. At this temperature, the r.m.s. speed of the atom $\sqrt{k_{\text{B}}T/m}$ is on the order of 0.02~nm$/$ns, which means that the atom moves by $\lesssim0.1$nm during the $5$-ns transient time staying at $5\text{s}5\text{p}~^{1}\text{P}_{1}$. To good approximation, the vibrational state of the atom remains during the transient at $5\text{s}5\text{p}~^{1}\text{P}_{1}$. 

Third, the fast spontaneous decay rate of the state $5\text{s}5\text{p}~^{1}\text{P}_{1}$ causes an incoherent transition 
\begin{eqnarray} 
|[5\text{s}5\text{p}~^{1}\text{P}_{1}]m_J=-1,m_I\rangle\otimes |n\rangle \rightsquigarrow
5\text{s}^2~^{1}\text{S}_{0}|m_I\rangle\otimes |n\rangle,\nonumber\\
\label{step3}
\end{eqnarray}
which returns the state back to the ground state. The transitions in Eqs.~(\ref{step1}),~(\ref{step2}), and~(\ref{step3}) involve a state with a common $m_I$, and similar transitions can happen with a superposition state of different $m_I$-eigenstates. So, the vibrational quantum number is lowered by one following Eqs.~(\ref{step1}),~(\ref{step2}), and~(\ref{step3}). As along as the atom is in a state with the vibrational quantum number $n$ larger than zero, the three-step cooling can proceed following Eqs.~(\ref{step1}),~(\ref{step2}), and~(\ref{step3}) until the atom reaches the ground.

\section{nuclear-spin-preserving cooling }\label{sec03}

\subsection{Hyperfine interaction mixes the nuclear spins}
The hyperfine interaction in $5\text{s}5\text{p}~^{1}\text{P}_{1}$ is not considered above. In practice, hyperfine interaction causes Eq.~(\ref{step2}) to become
\begin{eqnarray} 
|[5\text{s}5\text{p}~^{3}\text{P}_{0}]m_F=m_I\rangle\otimes |n\rangle  \rightarrow
|[5\text{s}5\text{p}~^{1}\text{P}_{1}]m_F-1\rangle\otimes |n\rangle,\nonumber\\\label{step2-2}
\end{eqnarray}
where $|[5\text{s}5\text{p}~^{1}\text{P}_{1}]m_F-1\rangle\otimes |n\rangle$ is a hyperfine eigenstate that mixes nuclear spin states with $m_I, m_I\pm1, m_I\pm2$, where the mixing coefficients are determined by the detail of the hyperfine interaction. To understand this coupling, we note that in the presence of a magnetic field $B\mathbf{z}$, the Hamiltonian including the hyperfine interaction between the valence electrons and the nuclear spin is described by 
\begin{eqnarray} 
\hat{H}_{\text{hf}}&=&A\hat{\mathbf{I}}\cdot \hat{\mathbf{J}}+ Q\frac{ 3(\hat{\mathbf{I}}\cdot \hat{\mathbf{J}})^2+ 1.5\hat{\mathbf{I}}\cdot \hat{\mathbf{J}} -IJ(I+1)(J+1) }{2IJ(2I-1)(2J-1)}\nonumber\\
&& + g_J\mu_{\text{B}} \hat{\mathbf{J}}\cdot B\mathbf{z} -g_I\mu_{\text{n}} \hat{\mathbf{I}}\cdot B\mathbf{z}. \label{hyperfine01}
\end{eqnarray}
Here, $A$ and $Q$ are the nuclear magnetic dipole and electric quadrupole interaction constants, respectively, $\hat{\mathbf{I}}$ and $\hat{\mathbf{J}}$ are the nuclear spin and electron orbital angular momentum operators~(divided by the reduced Planck constant), respectively, $g_J$ and $g_I$ are the electron and nuclear $g$-factors, respectively~\cite{DASteck}, and $\mu_{\text{B}}$ and $\mu_{\text{n}}$ are the Bohr magneton and the nuclear magnetic moment, respectively. According to the measurement in Ref.~\cite{Kluge1974}, the hyperfine constants are $(A, Q)/2\pi=(-3.4,~39)$~MHz for $5\text{s}5\text{p}~^{1}\text{P}_{1}$, and the measured value for $\mu_{\text{n}}$ reported in Ref.~\cite{Olschewski1972} is $-1.0924\mu_{\text{N}}$, where $\mu_{\text{N}}$ is the nuclear magneton.

The hyperfine interaction in Eq.~(\ref{hyperfine01}) couples states with equal $m_J+m_I$, as shown in Appendix~\ref{app-B} so that there can be decoherence in the nuclear spin state during the cooling if we do not introduce extra schemes. To uncouple the electron state and the nuclear spin state, strong magnetic fields about 10~mT can be used so that the nuclear-spin coherence can be preserved with a 99$\%$ fidelity when qubits are defined with nuclear spin projections $\pm m_I$ as studied in Ref.~\cite{Reichenbach2007}.

In our theory, nuclear spin qubits are defined with $m_I=-I$ and $1-I$, so that the state $|[5\text{s}5\text{p}~^{1}\text{P}_{1}]-1, -I\rangle\otimes |n\rangle$ can't be coupled to another state by hyperfine interaction, but $|[5\text{s}5\text{p}~^{1}\text{P}_{1}]-1, 1-I\rangle\otimes |n\rangle$ is coupled with $|[5\text{s}5\text{p}~^{1}\text{P}_{1}]0, -I\rangle\otimes |n\rangle$. To remove the nuclear spin mixing by the hyperfine interaction, we propose to use $\pi$-polarized laser excitation of an electric dipole transition between $5\text{s}5\text{p}~^{1}\text{P}_{1} $ and a nearby $5\text{s}n\text{s}~^{1}\text{S}_{0}$ state. An electric dipole transition directly couples two states with the change of $m_J$ equal to that of the angular momentum of the photon of the laser field. As a result, $5\text{s}n\text{s}~^{1}\text{S}_{0}$ can be coupled with the state $|[5\text{s}5\text{p}~^{1}\text{P}_{1}]0, -I\rangle$, but can be coupled with neither $|[5\text{s}5\text{p}~^{1}\text{P}_{1}]-1, 1-I\rangle$ nor $|[5\text{s}5\text{p}~^{1}\text{P}_{1}]-1, -I\rangle$. When this coupling is strong, a large AC Stark shift can arise in $|[5\text{s}5\text{p}~^{1}\text{P}_{1}]0, -I\rangle$. When the AC Stark shift is large compared to the hyperfine interaction, the hyperfine-interaction-induced state mixing between $|[5\text{s}5\text{p}~^{1}\text{P}_{1}]-1, 1-I\rangle$ and $|[5\text{s}5\text{p}~^{1}\text{P}_{1}]0, -I\rangle$ is suppressed. The questions is, is there a $5\text{s}n\text{s}~^{1}\text{S}_{0}$ state sufficient near to $5\text{s}5\text{p}~^{1}\text{P}_{1} $ so that a large electric dipole matrix element exists?

\subsection{The method }\label{sec03A}
The suppression of the hyperfine interaction by AC Stark shifts of laser excitation requires that a state near to $5\text{s}5\text{p}~^{1}\text{P}_{1}$ should have a large electric dipole transition matrix element, so that a large Rabi frequency can arise for the transition. There are several candidates satisfying this condition, among which $5s4d~^1D_2$ and $5s6s~^1S_0$ are nearest. The reduced dipole matrix element between $5s5p~^1P_1$ and $5s4d~^1D_2$ is $1.92ea_0$~\cite{Cooper2018}, where $e$ is the elementary charge and $a_0$ is the Bohr radius; however, the transition between $5s5p~^1P_1$ and $5s4d~^1D_2$ requires a 6.5~$\mu$m~\cite{Xu2003} laser, which can be challenging since lasers with such wavelength may not be immediately available~\cite{Carrig2011,Rosenfeld2020}. 
On the other hand, the transition from $5s5p~^1P_1$ to $5s6s~^1S_0$ has a wavelength $1124$~nm~\cite{Rubbmark1978} for which a laser is readily available. The reduced dipole matrix element for this transition is about $2.09ea_0$~(see Appendix~\ref{app-A}), and with a laser of field intensity about $17$W$/$cm$^2$, a Rabi frequency $\Omega/2\pi=300$~MHz can be achieved for the transition between $|[5\text{s}5\text{p}~^{1}\text{P}_{1}]0, -I\rangle$ and $|[5\text{s}6\text{s}~^{1}\text{S}_{-}]0, -I\rangle$ as shown in Appendix~\ref{app-A}. By the AC Stark shift, the hyperfine coupling between  
$|[5\text{s}5\text{p}~^{1}\text{P}_{1}]-1, 1-I\rangle$ and $|[5\text{s}5\text{p}~^{1}\text{P}_{1}]0, -I\rangle$ is suppressed, resulting in the suppression of the polarization resolution in the spontaneous emission from $5\text{s}5\text{p}~^{1}\text{P}_{1}$ to the ground state. 

The spontaneous decay from the states $|[5\text{s}5\text{p}~^{1}\text{P}_{1}]-1, 1-I\rangle$ and $|[5\text{s}5\text{p}~^{1}\text{P}_{1}]-1, -I\rangle$ back to the ground state should not be frequency resolved so as to preserve the coherence of the nuclear spin qubit. In a B-field of Gauss scale, the Zeeman shift between the two nuclear spin states is negligible. However, there is still a MHz-scale energy difference between $|[5\text{s}5\text{p}~^{1}\text{P}_{1}]-1, 1-I\rangle$ and $|[5\text{s}5\text{p}~^{1}\text{P}_{1}]-1, -I\rangle$ due to the diagonal hyperfine interaction~(see Appendix~\ref{app-B}). To remove this energy difference, a highly detuned laser field of wavelength 424~nm for the transition between   
$5\text{s}5\text{p}~^{1}\text{P}_{1}$ and $5\text{s}15\text{d}~^{1}\text{D}_{2}$ can be employed. We choose the state $5\text{s}15\text{d}~^{1}\text{D}_{2}$ for it has a strong hyperfine interaction~\cite{PhysRevA.47.4725}, so that when the laser is tuned near to one of its $F$ states, $|[5\text{s}5\text{p}~^{1}\text{P}_{1}]-1, 1-I\rangle$ and $|[5\text{s}5\text{p}~^{1}\text{P}_{1}]-1, -I\rangle$ can obtain different AC Stark shifts due to the different coupling strengths determined by the angular momentum selection rule. As shown in Appendix~\ref{app-C} with data from Ref.~\cite{PhysRevA.47.4725}, the two states with $F=I+2$ and $F=I+1$ of $5\text{s}15\text{d}~^{1}\text{D}_{2}$ are separated by about $2\pi\times1.3$~GHz, so that when a left-hand polarized laser field is tuned near to, e.g., the $F=I+1$ state, the AC Stark shift for the state $|[5\text{s}5\text{p}~^{1}\text{P}_{1}]-1, -I\rangle$ is negligible for it can only couple with the $F=I+2$ state, while AC Stark shifts can readily appear for other relevant $5\text{s}5\text{p}~^{1}\text{P}_{1}$ states. As a result, the frequency resolution in the spontaneous emission from $5\text{s}5\text{p}~^{1}\text{P}_{1}$ to the ground state can be avoided.

Compared to the theory of Ref.~\cite{Reichenbach2007} which needs a magnetic field over 10~mT, the method of cooling here is with a B-field on the order of 1~G. The sideband cooling with a low B-field is compatible with setups used in recent experiments with nuclear-spin qubits, where a B-field equal to $11$~G~\cite{Barnes2022}, $4.11$~G~\cite{Ma2022}, or a value in the range $(0,~18]$~G~\cite{Jenkins2022} was used with $^{87}$Sr~\cite{Barnes2022} or $^{171}$Yb~\cite{Ma2022,Jenkins2022}.

\section{Master equation analysis of cooling dynamics}\label{sec04}
We consider a nuclear spin qubit defined by the two maximal spin projections along the quantization axis, namely, the two lowest in energy~\cite{Barnes2022}, $\lvert\uparrow\rangle=|m_I=1-I\rangle$, and $\lvert\downarrow\rangle=|m_I=-I\rangle$; this type of qubit can be initialized with a bias magnetic field, while qubits defined with other nuclear spin states require extra fields~\cite{Omanakuttan2021}. To simplify the notation, we label the intrinsic state of, e.g., the ground-state atom, by $|[5s^2~^1S_0]0, \uparrow(\downarrow)\rangle$, where $0$ denotes the value of $m_J$ for the state and the arrow denotes the nuclear spin. With $\pi$ polarized laser fields employed for the atomic transitions, a general qubit state in the ground state
\begin{eqnarray}
(\alpha|[5s^2~^1S_0]0, \uparrow\rangle+ \beta|[5s^2~^1S_0]0, \downarrow\rangle)\otimes |n+1\rangle ,\nonumber
\end{eqnarray}
is excited to 
\begin{eqnarray}
(\alpha|[5s5p~^3P_0]0, \uparrow\rangle+ \beta|[5s5p~^3P_0]0, \downarrow\rangle)\otimes |n\rangle ,\label{3p0state}
\end{eqnarray}
where $|\alpha|^2+|\beta|^2=1$ and one vibrational quantum is removed in the above transition. In principle, there can be a change of the relative phase between the two spin components in the above transition, which can be amended by first exciting the state from the ground to the clock states, applying single-qubit phase gate to nuclear-spin qubits in the clock state, and then exciting the clock state to the $^1P_1$ state; an alternative is to design a compensating relative phase in the Rabi frequencies for the two nuclear spin states in the Raman transition between the clock and the $^1P_1$ states. Via an intermediate state with a two-photon $\sigma^-$ polarized transition, the state in Eq.~(\ref{3p0state}) is excited to 
\begin{eqnarray}
&&\Big[\alpha(\alpha_0|[5s5p~^1P_1]0, \downarrow\rangle+ \alpha_-|[5s5p~^1P_1]-1, \uparrow\rangle) \nonumber\\
&& +\beta |[5s5p~^1P_1]-1, \downarrow\rangle
 \Big]\otimes |n\rangle , \label{alpha0}
\end{eqnarray}
where the coefficients $\alpha_0,\alpha_-$ are determined by the hyperfine interaction that couples states with equal $m_J+m_I$~(see Appendix~\ref{app-B}). We note that Eq.~(\ref{alpha0}) is shown for illustration; in practice, Eq.~(\ref{alpha0}) is split into several different states with different energies in the presence of hyperfine interaction. 

To shift away the transitions with the nuclear spin flip in Eq.~(\ref{alpha0}), a $\pi$-polarized laser field is used to excite $5s5p~^1P_1$ to $5s6s~^1S_0$, where the angular momentum selection rule allows the transition between $|[5s5p~^1P_1]0, \downarrow\rangle$ and $|[5s6s~^1S_0]0, \downarrow\rangle$, while the other two state components in Eq.~(\ref{alpha0}) are not excited. As a result, the state component $|[5s5p~^1P_1]0, \downarrow\rangle$ can obtain an AC Stark shift large compared to the hyperfine interaction, leading to suppression of the hyperfine interaction, i.e., $\alpha_0\rightarrow0$ in Eq.~(\ref{alpha0}). 

Once the nuclear-spin flip state $|[5s5p~^1P_1]0, \downarrow\rangle$ is suppressed in Eq.~(\ref{alpha0}), polarization resolution in the spontaneous emission is suppressed. However, there is a MHz-scale energy difference between $|[5s5p~^1P_1]-1, \downarrow\rangle$ and $|[5s5p~^1P_1]-1, \uparrow\rangle$ mainly from the diagonal hyperfine interaction, leading to frequency resolution of the spontaneous emission. To remove this energy difference, one can excite $5s5p~^1P_1$ to a certain hyperfine substate of a $^1D_2$ state. A useful $^1D_2$ state for this purpose shall possess a large hyperfine interaction, ensuring different $F$ states being well separated. Then, tuning the frequency of a $\sigma^-$-polarized laser near to, e.g., the $F=I+1$ substate of $^1D_2$, the state $|[5s5p~^1P_1]-1, \downarrow\rangle$ barely acquires any AC Stark shift for it can't be excited to the $F=I+1$ state with the $\sigma^-$ laser field, while $|[5s5p~^1P_1]-1, \uparrow\rangle$ can acquire an AC Stark shift to compensate the energy difference between it and $|[5s5p~^1P_1]-1, \downarrow\rangle$. As discussed in Sec.~\ref{sec03}, we choose 5s15d~$^1$D$_2$ for it has a large hyperfine interaction.

In the cooling process, the excitation from the ground to the clock state can be achieved with high accuracy because of the long lifetime of the clock state, so that we analyze the cooling fidelity by starting from a state like Eq.~(\ref{3p0state}). In particular, we would like to see how precise we can map the state in Eq.~(\ref{3p0state}) to 
\begin{eqnarray}
(\alpha|[5s^2~^1S_0]0, \uparrow\rangle+ \beta| ][5s^2~^1S_0]0, \downarrow\rangle)\otimes |n\rangle ,\nonumber
\end{eqnarray}
during which the vibrational quantum number does not change, so that we can omit it when writing the state vectors in the analysis. The time dynamics is described by 
\begin{eqnarray}
 \frac{d\hat{\rho}}{dt} &=& i (\hat{\rho}\hat{H}- \hat{H} \hat{\rho}) +\sum_{i=0}^{8}\left[2\hat{c}_i\hat{\rho}\hat{c}_i^\dag- \hat{c}_i^\dag\hat{c}_i \hat{\rho}-\hat{\rho}\hat{c}_i^\dag\hat{c}_i  \right]/2.\nonumber\\
 \label{Lindblad1}
\end{eqnarray}
Here, $\hat{\rho}$ is the density matrix of the atomic state, and the Hamiltonian is
\begin{eqnarray}
 &&\hat{H}_{\text{hf}} + \bigg\{\frac{\Omega_{\text{eff}}}{2}\sum_{s_\text{z}\in\{\uparrow,\downarrow\}} |[5s5p~^1P_1]-1, s_\text{z}\rangle\langle [5s5p~^3P_0]0, s_\text{z}| \nonumber\\
 &&
 +  \frac{ \Omega_{\text{ps}}}{2}   |[5s6s~^1S_0]0, \downarrow\rangle\langle [5s5p~^1P_1]0, \downarrow|  +  \frac{ \Omega_{\text{pd}}}{2}\nonumber\\
 && \cdot \bigg( |[5s15d~^1D_2]F=\frac{13}{2}, m_F=-\frac{13}{2}\rangle\langle [5s5p~^1P_1]-1, \downarrow| \nonumber\\
 && +\xi_0|[5s15d~^1D_2]F=\frac{13}{2}, m_F=-\frac{11}{2}\rangle\langle [5s5p~^1P_1]0, \downarrow| \nonumber\\
 && +\xi_1|[5s15d~^1D_2]F=\frac{13}{2}, m_F=-\frac{11}{2}\rangle\langle [5s5p~^1P_1]-1, \uparrow| \nonumber\\
 && +\xi_2|[5s15d~^1D_2]F=\frac{11}{2}, m_F=-\frac{11}{2}\rangle\langle [5s5p~^1P_1]0, \downarrow| \nonumber\\
 && +\xi_3|[5s15d~^1D_2]F=\frac{11}{2}, m_F=-\frac{11}{2}\rangle\nonumber\\
 && ~~~  \langle [5s5p~^1P_1]-1, \uparrow|\bigg)+  \text{H.c.}\bigg\}\nonumber\\
 &&  + (\Delta+\Delta_{\text{pd}} )\sum_{k=0}^2 | S_k \rangle\langle S_k|  + \Delta \sum_{k=3}^6 | S_k \rangle\langle S_k|, \label{coolingH}
\end{eqnarray}
where $\hat{H}_{\text{hf}}$ is in a rotating frame derived from Eq.~(\ref{hyperfine01}), $\Omega_{\text{eff}}$ is the effective two-photon Rabi frequency between $5s5p~^1P_1$ and $5s5p~^3P_0$, $\Omega_{\text{ps}}$ is the $1124$~nm infrared-laser Rabi frequency between $5s5p~^1P_1$ and $5s6s~^1S_0$, $\Omega_{\text{pd}}$ is the $424$~nm UV-laser Rabi frequency between $5s5p~^1P_1$ and $5s15d~^1D_2$, the factors $\{\xi_j,j=0-3\}$ are angular momentum factors shown in Appendix~\ref{app-C}, $|S_k\rangle$ with $k=0-2$ are the three $5s15d$ states in the bracket $(\cdots)$ of Eq.~(\ref{coolingH}), $|S_k\rangle$ with $k=3-6$ are the states including $|[5s6s~^1S_0]0, \downarrow\rangle$ and the three $5s5p$ states in the bracket $(\cdots)$ of Eq.~(\ref{coolingH}), $\Delta_{\text{pd}}$ is the detuning for the dipole transition of the UV laser, and $\Delta$ is a detuning set to tune resonance for the transition between $|[5s5p~^1P_1]-1, s_\text{z}\rangle$ and $|[5s5p~^3P_0]0, s_\text{z}\rangle$; this latter detuning is added for when we use the UV laser field to induce AC Stark shifts in $|[5s5p~^1P_1]-1, s_\text{z}\rangle$, the states $|[5s5p~^1P_1]-1,\uparrow\rangle$ and $|[5s5p~^1P_1]-1,\downarrow\rangle$ can finally acquire a common, nonzero energy $-\Delta$ in the rotating frame. To simplify the numerical simulation, the hyperfine interaction in $5s15d~^1D_2$ is included with hyperfine eigenstates~[when laser detuning is accounted for, see Eq.~(\ref{app-Hmatrix})]. As analyzed in Appendix~\ref{app-C}, one can estimate by the measured hyperfine constants in Ref.~\cite{PhysRevA.47.4725} that the $F=\frac{13}{2}$ state of $5s15d~^1D_2$ is lower by about $2\pi\times1.3$~GHz than the $F=\frac{11}{2}$ state. 

In Eq.~(\ref{coolingH}), we ignore the coupling between $| [5s5p~^1P_1]1, \downarrow\rangle$ and $5s15d~^1D_2$ because $| [5s5p~^1P_1]1, \downarrow\rangle$ is populated neither directly nor indirectly~(via hyperfine interaction) from $5s5p~^3P_0$, though it can be populated via spontaneous emission from $5s6s~^1S_0$. However, $5s6s~^1S_0$ only has a negligible population when it is coupled to $|[5s5p~^1P_1]0,\downarrow\rangle$. In the present cooling scheme,  $|[5s5p~^1P_1]0,\downarrow\rangle$ is barely populated, least to say how negligible the population in $| [5s5p~^1P_1]1, \downarrow\rangle$ is via the higher-order process. This is why we can ignore the energy shift of $5s15d~^1D_2$ induced by the coupling between $| [5s5p~^1P_1]1, \downarrow\rangle$ and $5s15d~^1D_2$.

In Eq.~(\ref{hyperfine01}), the collapse operators for $5s5p~^1P_1$ are 
\begin{eqnarray} 
 \hat{c}_0 &=& \sqrt{\Gamma_\text{p}/3} \Big\{|[5s^2~^1S_0]0, \uparrow\rangle\langle  [5s5p~^1P_1]-1, \uparrow|\nonumber\\
  && +  |[5s^2~^1S_0]0, \downarrow\rangle\langle  [5s5p~^1P_1]-1, \downarrow|\Big\},\nonumber\\
 \hat{c}_1 &=&-\sqrt{\Gamma_\text{p}/3} |[5s^2~^1S_0]0, \downarrow\rangle\langle  [5s5p~^1P_1]0, \downarrow|,\nonumber\\
 \hat{c}_2 &=&\sqrt{\Gamma_\text{p}/3}  |[5s^2~^1S_0]0, \downarrow\rangle\langle  [5s5p~^1P_1]1, \downarrow| ,\label{collapseOp}
\end{eqnarray}
those for $5s6s~^1S_0$ are 
\begin{eqnarray} 
 \hat{c}_3 &=&\sqrt{\Gamma_\text{s}}| [5s5p~^1P_1]0, \downarrow\rangle\langle  [5s6s~^1S_0]0, \downarrow|,\nonumber\\ 
 \hat{c}_4 &=&\sqrt{\Gamma_\text{s}}| [5s5p~^1P_1]1, \downarrow\rangle\langle  [5s6s~^1S_0]0, \downarrow|,\nonumber\\ 
 \hat{c}_5 &=&\sqrt{\Gamma_\text{s}}| [5s5p~^1P_1]-1, \downarrow\rangle\langle  [5s6s~^1S_0]0, \downarrow|,\label{collapseSP}
\end{eqnarray}
and those for the $5s15d~^1D_2$ states are 
\begin{eqnarray} 
 \hat{c}_k &=& \sqrt{\Gamma_\text{d}} |\mathscr{A} \rangle\langle S_k| ,   \label{collapseOd}
\end{eqnarray}
where $k=6-8$ and $|S_k\rangle$ runs through the three $|[5s15d~^1D_2]F, m_F\rangle$ states in Eq.~(\ref{coolingH}) and $|\mathscr{A}\rangle$ is a virtual reservoir state that does not respond to the laser excitation. Here, $\Gamma_\text{p}/2\pi=32$~MHz~\cite{Xu2003,Millen2010}, $\Gamma_\text{s}/2\pi=3.0$~MHz~\cite{PhysRevA.46.1248}, and $\Gamma_\text{d}/2\pi=0.47$~MHz~\cite{PhysRevA.52.4416}. Note that in principle the decay rates in $\hat{c}_k$ with $k\in[6,8]$ should be smaller than the linewidth of the state $5s15d~^1D_2$ due to angular momentum selection rules, but a larger decay rate is employed so as to give a lower bound for the cooling fidelity.

\begin{figure}
\includegraphics[width=3.0in]
{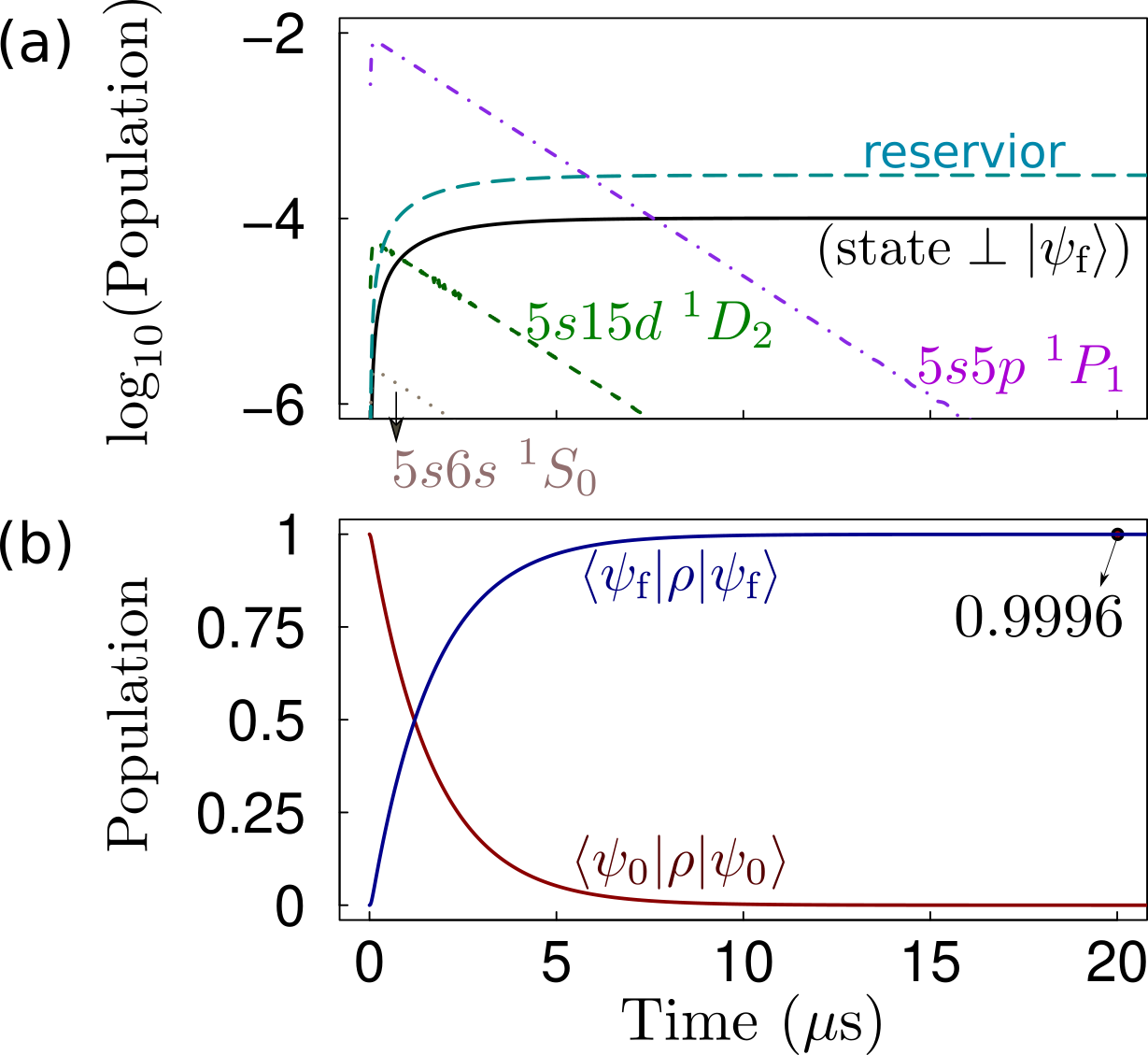}
\caption{State dynamics in the cooling simulated by the master equation in Eq.~(\ref{Lindblad1}) with $(\Omega_{\text{eff}}, \Omega_{\text{pd}},\Omega_{\text{ps}}, \Delta_{\text{pd}}, \Delta, \Gamma_{\text{p}},\Gamma_{\text{s}}, \Gamma_{\text{d}})/2\pi=(1, 144.27,300,-1700, 3.8826,32, 3,0.47)$~MHz, $B=1$~G, and initial state $|\psi_0\rangle$. The hyperfine constants $(A, Q)/2\pi$ are $(-3.4,~39)$~MHz for $5\text{s}5\text{p}~^{1}\text{P}_{1}$ and $(-194,-75)~$MHz for $5\text{s}15\text{d}~^{1}\text{D}_{2}$. (a) The solid, long-dashed, dash-dotted, short-dashed, and dotted curves show the populations of log10 scale in the states $\lvert \perp\rangle=(|[5s^2~^1S_0]0, \uparrow\rangle-|[5s^2~^1S_0]0, \downarrow\rangle)\otimes |n\rangle/\sqrt{2} $, $\lvert\mathscr{A}\rangle$, $5s5P~^1P_1$, $5s15d~^1D_2$, and $5s6s~^1S_0$, respectively. $\lvert \perp\rangle$ is the state perpendicular to the final target state $|\psi_{\text{f}}\rangle=(|[5s^2~^1S_0]0, \uparrow\rangle+  |[5s^2~^1S_0]0, \downarrow\rangle)\otimes |n\rangle/\sqrt{2} $. The final populations in $\lvert \perp\rangle$ and $\lvert\mathscr{A}\rangle$ are $1.0\times10^{-4}$ and $2.9\times10^{-4}$, respectively. (b) Evolution of the population of the states $|\psi_0\rangle=(|[5s5p~^3P_0]0, \uparrow\rangle+  |[5s5p~^3P_0]0, \downarrow\rangle)\otimes |n\rangle/\sqrt{2} $ and $|\psi_{\text{f}}\rangle$. The populations in $|\psi_0\rangle$ and $|\psi_{\text{f}}\rangle$ at $20~\mu$s are $7\times10^{-6}$ and $0.9996$, respectively, and the main population loss is in the reservoir state and the state perpendicular to $|\psi_{\text{f}}\rangle$ as shown in (a).  \label{figure3} }
\end{figure}

The laser parameters are chosen with the following considerations. (i) First, the effective Rabi frequency $\Omega_{\text{eff}}$ for the transition between the clock state and $5s5p~^1P_1$ is via a highly off-resonant intermediate state, so that it is in general small and we use $\Omega_{\text{eff}}/2\pi=1$~MHz in the numerical example. (ii) Second, the $\pi$-polarized laser for the transition between $5s5p~^1P_1$ and $5s6s~^1S_0$ is relevant for $| [5s5p~^1P_1]0, \downarrow\rangle$. $| [5s5p~^1P_1]0, \downarrow\rangle$ can be populated via hyperfine interaction, and our purpose is to suppress its population via the AC stark effect. To induce a large shift, the laser is tuned resonant for the transition between $| [5s5p~^1P_1]0, \downarrow\rangle$ and $| [5s6s~^1S_0]0, \downarrow\rangle$, and the laser Rabi frequency $\Omega_{\text{ps}}$ shall be much larger than the hyperfine interaction. Therefore, we can set, as an example, $\Omega_{\text{ps}}/2\pi=300$~MHz, for which $| [5s5p~^1P_1]0, \downarrow\rangle$ can exhibit a shift $\Omega_{\text{ps}}/2$ that is much larger than the hyperfine interaction strength. Larger $\Omega_{\text{ps}}$ can work for the theory but requires stronger laser power. (iii) Third, the excitation of $5s15d~^1D_2$ is for balancing the energies of $| [5s5p~^1P_1]-1, \uparrow\rangle$ and
$| [5s5p~^1P_1]-1, \downarrow\rangle$, which have frequencies $2\pi\times (-10.05, -6.95)$~MHz mainly from the diagonal hyperfine interaction $2\pi\times (-8.65, -5.55)$~MHz in a B-field of 1~G. Because the AC Stark shift in a highly off-resonant dressing is $-\frac{1}{4}($Rabi$^2/$detuning$)$~\cite{Shi2018pra_m}, one can tune the $\sigma^-$-polarized laser to the blue side of the transition $| [5s5p~^1P_1]-1, \downarrow\rangle\rightarrow | [5s15d~^1D_2]F=13/2, m_F=-13/2\rangle$ with a detuning $\Delta_{\text{pd}}<0$. Because the $F=11/2$ state is higher than the $F=13/2$ state by about $E_{\text{hf}}=2\pi\times1.3$~GHz, the transition $| [5s5p~^1P_1]-1, \uparrow\rangle\rightarrow | [5s15d~^1D_2]F=11/2, m_F=-11/2\rangle$ has a detuning $\Delta_{\text{pd}}+E_{\text{hf}}$. For $\Delta_{\text{pd}}+E_{\text{hf}}<0$, the AC Stark shift of $| [5s5p~^1P_1]-1, \uparrow\rangle$ is also positive. The detuning of $| [5s5p~^1P_1]-1, \uparrow\rangle$ is smaller and it can obtain a larger shift compared to $| [5s5p~^1P_1]-1, \downarrow\rangle$, so that the energy difference between $| [5s5p~^1P_1]-1, \downarrow\rangle$ and $| [5s5p~^1P_1]-1, \uparrow\rangle$ can be compensated. In order to have negligible population in 5s15d, the dressing detuning shall be much larger than the dressing Rabi frequency. As an example, we choose $\Delta_{\text{pd}}/2\pi=-1.7$~GHz, i.e., about 400~MHz over the F=$11/2$ state. The value of $\Omega_{\text{pd}}$ can be searched for achieving the same energies for $| [5s5p~^1P_1]-1, \downarrow\rangle$ and $| [5s5p~^1P_1]-1, \uparrow\rangle$, which we numerically found $\Omega_{\text{pd}}/2\pi=144.27$~MHz; note that different $\Delta_{\text{pd}}$ will lead to different $\Omega_{\text{pd}}$. With these parameters, the final, common frequency $\nu$ for $| [5s5p~^1P_1]-1, \downarrow\rangle$ and $| [5s5p~^1P_1]-1, \uparrow\rangle$ is in general nonzero. We have $\nu=2\pi\times(-3.8826)$~MHz with the above parameters, which means that we shall have a detuning $-\nu$ in the two-photon transition between the clock and the $5s5p~^1P_1$ states so as to recover the resonant condition.       

By using Eq.~(\ref{Lindblad1}) with $(\Omega_{\text{eff}}, \Omega_{\text{pd}},\Omega_{\text{ps}}, \Delta_{\text{pd}}, \Delta)/2\pi=(1, 144.27,300,-1700, 3.8826)$~MHz and the decay rates shown around Eq.~(\ref{collapseOd}), we numerically simulated the time evolution of the system by QuTip~\cite{Johansson2012,Johansson2013}. We found that two eigenstates $|e_\uparrow\rangle$ and $|e_\downarrow\rangle$ of the Hamiltonian driven by the hyperfine interaction and laser excitation highly overlap with two pure spin states, namely, we found
\begin{eqnarray} 
\langle [5s5p~^1P_1]-1, \uparrow|e_\uparrow\rangle&=& 0.99409,\nonumber\\
\langle [5s5p~^1P_1]-1, \downarrow|e_\downarrow\rangle&=& 0.99910. \label{overlpa}
\end{eqnarray}
The states $|e_\uparrow\rangle$ and $|e_\downarrow\rangle$ have a common eigenenergy $2\pi\times(-3.8826)$~MHz, which is why we set $ \Delta/2\pi=3.8826$~MHz. Though $|e_\uparrow\rangle$ and $|e_\downarrow\rangle$ have populations in other state components, their decay rates are much smaller than that of $5s5p~^1P_1$, so that when the cooling starts, the spontaneous emission takes the atom back to the ground state with a high fidelity. 

\begin{table}[ht]
  \centering
  \begin{tabular}{|c|c|c|c|c|c|c|}
    \hline
    $\alpha/\beta$& $1/10$ & $1/3$& $1/2$ & 2 & 3&10
     \\\hline
    Fidelity&99.99$\%$& 99.99$\%$& 99.98$\%$&99.93$\%$ & 99.92$\%$ & 99.91$\%$  \\   
   \hline    
  \end{tabular}
  \caption{Fidelity of the cooling, $\langle \psi_{\text{f}} |\rho(t)|\psi_{\text{f}}\rangle $, at $t=20~\mu$s starting from the initial state $|\psi_0\rangle=(\alpha|[5s5p~^3P_0]0, \uparrow\rangle+  \beta|[5s5p~^3P_0]0, \downarrow\rangle)\otimes |n\rangle  $, where $|\psi_{\text{f}}\rangle=(\alpha|[5s^2~^1S_0]0, \uparrow\rangle+ \beta |[5s^2~^1S_0]0, \downarrow\rangle)\otimes |n\rangle $. The parameters used in the simulation are the same to those in Fig.~\ref{figure3}. \label{table1}  }
\end{table}

As discussed previously, the excitation from the ground state to the clock state can proceed with a high fidelity, so that we start from the initial state in the clock-state space, $|\psi_0\rangle=(|[5s5p~^3P_0]0, \uparrow\rangle+  |[5s5p~^3P_0]0, \downarrow\rangle)\otimes |n\rangle/\sqrt{2} $, from which a two-photon transition via an intermediate state can excite it to the
$5s5p~^1P_1$ state which rapidly decays back to the ground state. The desired final state is $|\psi_{\text{f}}\rangle=(|[5s^2~^1S_0]0, \uparrow\rangle+  |[5s^2~^1S_0]0, \downarrow\rangle)\otimes |n\rangle/\sqrt{2} $. As shown in the numerical result in Fig.~\ref{figure3}, the fidelity to cool the initial state $(|[5s^2~^1S_0]0, \uparrow\rangle+  |[5s^2~^1S_0]0, \downarrow\rangle)\otimes |n+1\rangle/\sqrt{2} $ to the desired final state $|\psi_{\text{f}}\rangle$ is about $99.96\%$. The final population loss is mainly in the reservoir state $\lvert\mathscr{A}\rangle$, and in the state $\lvert \perp\rangle=(|[5s^2~^1S_0]0, \uparrow\rangle-|[5s^2~^1S_0]0, \downarrow\rangle)\otimes |n\rangle/\sqrt{2} $. The cooling fidelity shows a weak dependence on the value of $|\alpha/\beta|$ in Eq.~(\ref{3p0state}). However, Table~\ref{table1} shows that the fidelity decreases slowly when $|\alpha/\beta|$ increases. This decrease is due to that the state $|e_\uparrow\rangle$ has a smaller overlap with the correct state component shown in Eq.~(\ref{overlpa}). However, we found that the fidelity is $99.909\%$ even with $|\alpha/\beta|=100$. This means that the theory can easily have a coherence-preserving cooling fidelity over $99.9\%$ in a weak magnetic field, and much higher fidelity can be achieved with stronger laser fields for suppressing the hyperfine interaction.

\section{influence from laser parameters}\label{Sec05}
\subsection{Fluctuation of Rabi frequency and detuning}\label{sec05A}
The fluctuation of the power and frequency of the lasers has a minor influence on the cooling. We don't discuss the transition from the ground to the clock state for it doesn't involve change of angular momentum and it can be realized with a high fidelity. Beside this laser, there are three sets of lasers, one for the transition $5s5p~^3P_0\rightarrow5s5p~^1P_1$, one for $5s5p~^1P_1\rightarrow5s6s~^1S_0$, and one for $5s5p~^1P_1\rightarrow5s15d~^1D_2$. Below, we discuss the influence on the cooling from the fluctuation of Rabi frequency and detuning of the lasers.

(i) The fluctuation of the Rabi frequency $\Omega_{\text{eff}}$ and detuning $\Delta$ for $5s5p~^3P_0\rightarrow5s5p~^1P_1$ can slow down the cooling but barely impacts the fidelity. The cooling can proceed when the population is transferred from $5s5p~^3P_0$ to $5s5p~^1P_1$. So, larger $\Omega_{\text{eff}}$ and the resonant condition $\Delta+\nu=0$ can facilitate the population transfer, while smaller $\Omega_{\text{eff}}$ or off-resonant condition can lead to slower cooling. For example, with $\Omega_{\text{eff}}/2\pi=2$~MHz while other parameters are the same as in Fig.~\ref{figure3}, simulation shows that the cooling can reach the final fidelity of Fig.~\ref{figure3} at a much earlier time $5~\mu$s. But with $\Delta=0$, i.e., off-resonant with a detuning $\nu$, while other parameters are the same as in Fig.~\ref{figure3}, simulation shows that the cooling reaches a fidelity $0.9991$ at $20~\mu$s, and the fidelity $0.9996$ is achieved at a later time $26~\mu$s. However, the above discussion is based on that the ratio of the Rabi frequencies for the two nuclear spin states does not change.

(ii) The cooling is not sensitive to small fluctuation of power and frequency of the laser for the transition $5s5p~^1P_1\rightarrow5s6s~^1S_0$. Because the $\pi$ polarization of the laser, only the state $| [5s5p~^1P_1]0, \downarrow\rangle$ is excited while the other two states $| [5s5p~^1P_1]-1, \uparrow\rangle$ and $| [5s5p~^1P_1]-1, \downarrow\rangle$ are not. The numerical example of Fig.~\ref{figure3} and Table~\ref{table1} assumed $\Omega_{\text{ps}}/2\pi=300$~MHz, a little deviation from this value alters the AC Stark shift, but as long as the shift is large compared to the hyperfine interaction, the hyperfine-interaction-induced spin mixing is suppressed. For example, with $\Omega_{\text{ps}}/2\pi=250$~MHz, i.e., smaller by $1/6$, while other parameters are the same as in Fig.~\ref{figure3}, we numerically found that a cooling fidelity $0.9996$ can still be achieved at $20~\mu$s. Likewise, by adding a dtuning, e.g.,  $2\pi\times10$~MHz, to this laser while preserving all the parameters as in Fig.~\ref{figure3}, a fidelity $0.9996$ is still achieved at $20~\mu$s.  

(iii) The fluctuation of the laser parameters for the transition $5s5p~^1P_1\rightarrow5s15d~^1D_2$ can influence the cooling fidelity. This is because when the Rabi frequency or detuning is not set so as to have the same frequency for the two states in Eq.~(\ref{overlpa}), decoherence will occur due to the frequency resolution. For one example, with $\Omega_{\text{pd}}/2\pi=140$~MHz while other parameters kept the same as used in Fig.~\ref{figure3}, the cooling fidelity is $0.99946$ at $20~\mu$s, and it can only reach $0.99947$ even at $30~\mu$s due to a relatively large population $2.6\times10^{-4}$ in $\lvert\perp\rangle$; the final population in $\lvert\perp\rangle$ is $1.0\times10^{-4}$ for the case simulated in Fig.~\ref{figure3}. For another example, with $\Delta_{\text{pd}}/2\pi=-1.75$~GHz while other parameters are the same as in Fig.~\ref{figure3}, the cooling fidelity is $0.9988$ at or beyond $20~\mu$s. This small cooling fidelity is due to that the two states in Eq.~(\ref{overlpa}) have a relatively large energy separation $2\pi\times0.42$~MHz which leads to a large final population $9.2\times10^{-4}$ in $\lvert\perp\rangle$. 

\subsection{Influence from laser polarization impurity}\label{sec05B}
The cooling depends on high purity in laser polarization. We follow Ref.~\cite{Chen2022} and use polarization intensity impurity $\chi$ for this discussion, where $\chi=0$ denotes perfect polarization. We discuss the laser polarization impurity of the three sets of laser fields discussed in Sec.~\ref{sec05A} about their influence on the cooling. 

(i) The transition $5s5p~^3P_0\rightarrow5s5p~^1P_1$ is a two-photon process via a highly detuned intermediate state. If each of the two lasers in the two-photon process has a polarization intensity impurity $\chi$, the effective Rabi frequency becomes $(1-\chi) \Omega_{\text{eff}}$ for the desired transitions $\lvert[5s5p~^3P_0]0,\uparrow(\downarrow)\rangle\rightarrow\lvert[5s5p~^1P_1]-1,\uparrow(\downarrow)\rangle$. That the effective Rabi frequency decreases can slow down the cooling as discussed in the last paragraph. But the wrong polarization can excite $\lvert[5s5p~^3P_0]0,\uparrow(\downarrow)\rangle$ to $\lvert[5s5p~^1P_1]m_J,m_I\rangle$ where $m_J\neq -1$ or $m_I$ is not equal to the correct value even when $m_J=-1$. There is some chance to accidentally lead to the correct state transition. For example, if the polarization should be $\sigma^-+\pi$ in the transition $5s5p~^3P_0 \rightarrow$intermediate$\rightarrow 5s5p~^1P_1 $, a wrong polarization $ \pi+\sigma^+$ in the two corresponding lasers can result in the correct state transfer. To estimate the worse case, we assume all polarization errors result in wrong state transfer, which means that there is a Rabi frequency $\chi\Omega_{\text{eff}}$ to create population loss to the correct ground state. The data in Fig.~\ref{figure3} shows that $\lvert\psi_{\text{f}}\rangle$ has a population over 0.5 beyond the time $2.4\pi/\Omega_{\text{eff}}$, and we can estimate that the total chance to have incorrect spontaneous emission due to wrong population in $5s5p~^1P_1 $ is around or below $\sin^2(2.4\pi\chi)$ when $\chi\ll1$. For $\chi=0.01$, it means that the cooling fidelity decreases by about $0.6\%$ due to the polarization impurity for this transition. 

(ii) The laser for $5s5p~^1P_1\rightarrow5s6s~^1S_0$ is assumed $\pi$ polarized. A wrong polarization with a small $\chi$ barely alters the AC Stark shift for shifting $\lvert[5s5p~^1P_1]0,\downarrow\rangle$, but can excite $\lvert[5s5p~^1P_1]-1,\uparrow(\downarrow)\rangle$ to $\lvert[5s6s~^1S_0]0,\uparrow(\downarrow)\rangle$. In the worst case when the wrong polarization is fully $\sigma^+$, the Rabi frequency is $-\sqrt{\chi}\Omega_{\text{ps}}$ for the transition $\lvert[5s5p~^1P_1]-1,m_I\rangle\rightarrow\lvert[5s6s~^1S_0]0,m_I\rangle$ where the minus sign is from the Clebsch-Gordan coefficient. This will result in common energy shifts to $\lvert[5s5p~^1P_1]-1,\uparrow(\downarrow)\rangle$ which does not hamper the cooling fidelity directly though it adds detuning to the two-photon excitation $5s5p~^3P_0 \rightarrow 5s5p~^1P_1 $ that will slow down the cooling as discussed above. However, when $-\sqrt{\chi}\Omega_{\text{ps}}$ is much larger than $\Omega_{\text{eff}}$, the transition $\lvert[5s5p~^1P_1]-1,m_I\rangle\rightarrow\lvert[5s6s~^1S_0]0,m_I\rangle$ can lead to half population in $5s5p~^1P_1$ while the other half in $5s6s~^1S_0$, and $5s6s~^1S_0$ decays with a rate $\Gamma_{\text{s}}$ that is about a tenth of the decay rate of $5s5p~^1P_1$. This means that the condition $|\sqrt{\chi}\Omega_{\text{ps}}/\Omega_{\text{eff}}|\gg1$ can reduce the cooling fidelity by about 0.1, which is significant. The condition $\chi<|\Omega_{\text{eff}}/\Omega_{\text{ps}}|^2$ can resolve this issue but it is difficult to realize since the parameters used in Sec.~\ref{sec04} require $\chi<10^{-5}$. A solution to this problem is to add a detuning to the laser field so that incorrect polarization in the field can only drive the transition $\lvert[5s5p~^1P_1]-1,m_I\rangle\rightarrow\lvert[5s6s~^1S_0]0,m_I\rangle$ with a detuning large compared to $|\sqrt{\chi}\Omega_{\text{ps}}|$. In this case, the state $\lvert[5s6s~^1S_0]0,m_I\rangle$ is barely populated from $\lvert[5s5p~^1P_1]-1,m_I\rangle$ and population loss from it can be avoided. 

(iii) The transition $5s5p~^1P_1\rightarrow5s15d~^1D_2$ is for balancing the energies of $\lvert[5s5p~^1P_1]-1,\uparrow(\downarrow)\rangle$. The detuning is $\Delta_{\text{pd}}$ when addressing $| [5s15d~^1D_2]F=13/2, m_F\rangle$ and $\Delta_{\text{pd}}+E_{\text{hf}}$ when addressing $| [5s15d~^1D_2]F=11/2, m_F\rangle$. The numerical example in Sec.~\ref{sec04} assumed $(\Omega_{\text{pd}}, \Delta_{\text{pd}})/2\pi=(144.27,~-1700)$~MHz where $E_{\text{hf}}/2\pi=1.3$~GHz. The laser is assumed $\sigma^-$ polarized, and polarization impurity can lead to dressing of the hyperfine state $F=9/2$ which is an extra state not included in the discussion of Sec.~\ref{sec04}. The $F=9/2$ state is above the $F=11/2$ state by about $2\pi\times1.1$~GHz shown in Appendix~\ref{app-C}, which means that it is unlikely to induce an extra AC Stark shift to hamper the cooling when $\sqrt{\chi}\xi_4\Omega_{\text{pd}}$ is much smaller than these detunings, where $\xi_4$ is a factor smaller than 1 defined similar to those in Eq.~(\ref{defineXi}). However, laser polarization impurity can reduce the Rabi frequency from $\xi_k\Omega_{\text{pd}}$ to $(1-\sqrt{\chi})\xi_k\Omega_{\text{pd}}$ for any of the desired dressings where $\xi_k$ with $k=0,1,2$, and $3$ are shown above Eq.~(\ref{defineXi}). The reduced Rabi frequencies can result in unbalanced energies of $\lvert[5s5p~^1P_1]-1,\uparrow(\downarrow)\rangle$, which will lead to reduced cooling fidelity as discussed in Sec.~\ref{sec05A}. Assuming $\chi=0.01$, the Rabi frequencies for dressing the target hyperfine states of $5s15d~^1D_2$ will decrease by $10\%$, and numerical simulation shows that the cooling fidelity is $0.9981$ at or beyond $20~\mu$s. With a worse polarization condition when $\chi=0.1$ which corresponds to a deduction of Rabi frequencies by $32\%$, the final cooling fidelity is about $0.9878$. 

\section{Discussions}\label{Sec06}
\subsection{Comparison with Ref.~\cite{Reichenbach2007}}
There are two differences between the present theory and that in Ref.~\cite{Reichenbach2007}. First, the hyperfine-interaction-induced spin mixing is suppressed by large Zeeman energy in Ref.~\cite{Reichenbach2007}, while here it is suppressed by AC Stark shift of laser fields. Second, the theory of Ref.~\cite{Reichenbach2007} depends on defining qubits with $\pm m_I$, while the present theory depends on defining qubits with $m_I=1-I, -I$, namely, the two lowest nuclear-spin states in the ground state.

\subsection{Cooling nuclear spin qubits in other atoms}
It is useful to discuss the application of the present theory with other alkaline-earth elements like calcium and barium, and some alkaline-earth-like transition-metal elements that have similar relevant level structures. The theory hinges on suppression of the hyperfine-interaction-induced spin flip via exciting the lowest $^1P_1$ state to a nearby state with a large Rabi frequency $\Omega_{\text{ps}}$ which may not be available for all elements. 

\subsubsection{Ytterbium, calcium, and barium} 
Another widely studied candidate for nuclear-spin quantum memories with neutral atoms is $^{171}$Yb. There is a relatively large hyperfine interaction in $6s6p~^1P_1$ of $^{171}$Yb, with $A/2\pi=-213$~MHz~\cite{Berends1992}~(the quadrupole interaction is zero for $I$ is $1/2$ with $^{171}$Yb). To use the present theory with $^{171}$Yb, a large AC Stark shift is required to suppress the hyperfine interaction, so it is not practical to use the present theory for cooling $^{171}$Yb. For example, Eq.~(\ref{overlpa}) shows that for $^{87}$Sr, a Rabi frequency $\Omega_{\text{ps}}/2\pi$ of about 300~MHz can result in an overlap over 0.99 between the state used in the cooling and the correct state; for $^{171}$Yb, we find that to reach a similar overlap over 0.99, a corresponding Rabi frequency over $2\pi\times$2.15~GHz should be available, which may be challenging. For $^{173}$Yb that has $I=5/2$, the hyperfine splitting in the lowest $^1P_1$ state is stronger than that of $^{171}$Yb~\cite{Deilamian1993} and is characterized with $(A,~Q)/2\pi\approx(60,600)$~MHz~\cite{Berends1992}. To realize an overlap over 0.99 between the state used in the cooling and the correct state as in Eq.~(\ref{overlpa}), we find that the strong hyperfine interaction in $^{173}$Yb requires $\Omega_{\text{ps}}>2\pi\times$5.4~GHz which is unlikely to be realizable. For this reason, we conclude that the present theory is applicable for AEL isotopes where the lowest $^1P_1$ state has a small enough hyperfine interaction. 

The present cooling scheme can work for $^{41}$Ca and $^{43}$Ca which have $I=7/2$ and a level structure that is compatible with the cooling theory. The ground state of calcium is $4s^2~^1S_0$ and the lowest $^1P_1$ state is $4s4p~^1P_1$. The stable calcium isotope $^{43}$Ca has a relatively weak hyperfine interaction with $(A,~Q)/2\pi=-(15.46,9.7)$~MHz~\cite{PhysRevC.26.2194}. For the radionuclide odd calcium isotope that can be assumed stable in quantum optics~(its half-life is about $10^5$ years), $^{41}$Ca, the $4s4p~^1P_1$ state has $(A,~Q)/2\pi=-(18.84,9.2)$~MHz~\cite{PhysRevC.26.2194}. The transition from $4s4p~^1P_1$ to $4s5s~^1S_0$ has a wavelength 1034.66~nm~\cite{Dammalapati2010} which is close to that in the case of $^{171}$Yb as shown in Fig.~\ref{figure1}, and this transition has a rate $2.435\times10^7s^{-1}$~\cite{Dammalapati2010} that is larger than the corresponding value shown in Eq.~(\ref{Weisskopf2}) used in the numerical example studied in this paper. This means that it should be easier to realize a large $\Omega_{\text{ps}}$ for calcium, and it is possible to suppress the hyperfine-interaction-induced spin mixing in calcium which is crucial for the theory to work. We find that the minimal $\Omega_{\text{ps}}/2\pi$ to realize an overlap over 0.99 between the state used in the cooling and the correct state as in Eq.~(\ref{overlpa}) shall be at least 490 and 580~MHz for $^{43}$Ca and $^{41}$Ca, respectively. $^{41}$Ca and $^{43}$Ca were not well studied as $^{87}$Sr~\cite{Kramida2020}, but according to the discussion in Appendix~\ref{app-A}, the numerical example shown in Figs.~\ref{figure1} can in principle be realized with $\Omega_{\text{ps}}$ up to $2\pi\times1$~GHz, which means that the present theory can work with $^{41}$Ca and $^{43}$Ca since they possess even larger transition rate in the infrared-laser transition for suppressing nuclear spin mixing. 

The two stable odd barium isotopes $^{135}$Ba and $^{137}$Ba have $I=3/2$ and the electronic ground and optical clock states have similar configurations to those of $^{171}$Yb~\cite{PhysRevA.47.R2446,PhysRevA.49.1158}. The spectra reported in Ref.~\cite{PhysRevA.49.1158} show that the hyperfine interaction in $6s6p~^1P_1$ leads to an frequency separation of 400.5~MHz and 457.2~MHz between the $F=5/2$ and $F=1/2$ levels for $^{135}$Ba and $^{137}$Ba, respectively. Comparing to $5s5p~^1P_1$ in $^{87}$Sr, the frequency separations of two nearby F levels in $6s6p~^1P_1$ of $^{135}$Ba and $^{137}$Ba are roughly four times larger. We suppose that the value of $\Omega_{\text{ps}}$ for achieving suppression of nuclear-spin mixing as in Eq.~(\ref{overlpa}) should be at least four times larger than that used in the example shown in Sec.~\ref{sec04}. However, the condition in Eq.~(\ref{overlpa}) is for a cooling fidelity over 0.999 with $^{87}$Sr shown in Table~\ref{table1}. for a rough estimate, we find that if the values of $A$ and $Q$ used in Sec.~\ref{sec04} are increased by four times while other parameters are the same, the two numbers on the right sides of the two lines of Eq.~(\ref{overlpa}) become about $0.984$ and $0.999$, respectively, which suggests that even with an infrared laser field of similar strength as used in Sec.~\ref{sec04}, a cooling fidelity of around 0.99 should be achievable with $^{135}$Ba and $^{137}$Ba.

\subsubsection{Zinc, cadmium, and mercury} 
There are some alkaline-earth-like transition-metal elements with nuclear spins and low-lying states similar to the elements discussed above. For example, the ground-clock transition has a wavelength 309, 332, and 266~nm for zinc, cadmium, and mercury, respectively~\cite{Garstang:62}, and it is possible to achieve high power UV laser fields for driving these transitions~\cite{Manzoor:22}. 

For zinc, the stable odd isotope $^{67}$Zn has $I=5/2$, and its lowest $^1P_1$ state, $4s4p~^1P_1$, has both a fast decay rate with a lifetime around $1.3$~ns~\cite{PhysRev.134.A56,Martinson1979,Chi2014} and a relatively weak hyperfine interaction with $(A,Q)/2\pi\approx(17.7,20.0)$~MHz~\cite{Kowalski1976}. We note that Eq.~(\ref{overlpa}) in the case of $^{87}$Sr shows that a Rabi frequency $\Omega_{\text{pd}}/2\pi$ of about 300~MHz is quite useful; here, we find that to reach a wavefunction overlap over 0.99 in an equation similar to Eq.~(\ref{overlpa}) for the case of $^{67}$Zn, a Rabi frequency about $2\pi\times$535~MHz is sufficient. Though we didn't find data for a strong transition between $4s4p~^1P_1$ and a higher state, the data in Ref.~\cite{Chi2014} for the triplet states $4s4p^3P_x$ with $x=0,1,2$ indirectly suggest that it is possible to have a strong transition for the singlet state as well. This suggests that the cooling theory can in principle be applied with $^{67}$Zn due to the weak hyperfine interaction.  

Both of the two stable odd cadmium isotopes $^{111}$Cd and $^{113}$Cd have a simple nuclear spin state with $I=1/2$ and they are recognized as a useful candidate for optical lattice clocks~\cite{Yamaguchi2019}. However, the hyperfine interaction in $5s5p~^1P_1$ of $^{111}$Cd and $^{113}$Cd is strong~\cite{PhysRev.178.153}, with $|A|/2\pi$ equal to about 150 and 240~MHz, respectively~\cite{PhysRev.134.A608}, which is comparable to that of $6s6p~^1P_1$ in $^{171}$Yb. We didn't find data about dipole matrix elements between $5s5p~^1P_1$ and a nearby state for suppressing the hyperfine-interaction-induced spin mixing, but that the decay rate of $5s5p~^1P_1$ in $^{111}$Cd and $^{113}$Cd~\cite{Yamaguchi2019} is more than three times of that in $^{171}$Yb suggests that the dipole matrix element between $5s5p~^1P_1$ and a nearby state in Cd is likely to be much larger. So, it is possible to achieve a much larger $\Omega_{\text{ps}}$ with reasonable laser powers for cooling $^{111}$Cd and $^{113}$Cd and we think that it might be possible to use the present theory with cadmium.  

Mercury is among the heaviest elements that were optically trapped for precision physics~\cite{PhysRevLett.100.053001,McFerran2012,PhysRevLett.114.230801}. The stable odd isotopes $^{199}$Hg and $^{201}$Hg have $I=1/2$ and $3/2$, respectively. However, the hyperfine interaction in the lowest $^1P_1$ state, $6s6p~^1P_1$, is so strong that the frequency separation between the $F=1/2$ and $F=3/2~(5/2)$ states is about $5$~GHz for $^{199}$Hg~($^{201}$Hg)~\cite{PhysRevA.33.2003,Leboucher1974} which suggests that it is challenging to apply the present cooling theory for mercury. Nonetheless, the lifetime of the $6s6p~^1P_1$ state in mercury is 1.31~ns~\cite{PhysRev.140.A1505} which is a quarter of that of the $6s6p~^1P_1$ state in $^{171}$Yb. This suggests that the dipole matrix element between $6s6p~^1P_1$ and a nearby state in $^{199}$Hg and $^{201}$Hg can be much larger than that in $^{171}$Yb, and it is difficult to say that the present cooling theory can't be used with mercury.    

The above discussions show that the theory shown with strontium as an example in this paper can be used with zinc and calcium with a high cooling fidelity. It may also be used with barium and cadmium but the cooling fidelity may not be high unless strong laser fields are available for suppressing the hyperfine interactions. We only studied AEL atoms in this paper and it is a question whether the theory can be extended to quantum control over nuclear spins of noble gas~\cite{Dantan2005,Katz2020,Shaham2022}.

\section{Conclusion}\label{Sec07}
We present a theory to cool $^{87}$Sr atoms with resolved sideband excitation from the ground state to the clock state quenched by two-photon excitation between the clock state and the fast-decaying $5s5p~^1P_1$ state. The nuclear-spin-changing process induced by the hyperfine interaction in $5s5p~^1P_1$ is suppressed by using laser excitation between 
$5s5p~^1P_1$ and nearby states. The suppression is achieved via the $m_J$-dependent AC Stark shift that is large compared to the hyperfine interaction. Numerical simulations with reasonable parameters indicate that a cooling fidelity over $99.9\%$ can be easily achieved with $^{87}$Sr. The cooling is not sensitive to fluctuation of intensities and frequencies of the lasers, but depends on high polarization purity in the laser fields. The theory can be used with some other alkaline-earth-like species like calcium, zinc, and barium.

\section*{ACKNOWLEDGMENTS}
The author thanks T. A. B. Kennedy for valuable inputs during the initial stage of this work, and thanks Yan Lu for helpful discussions. This work is supported by the National Natural Science Foundation of China under Grants No. 12074300 and No. 11805146, the Innovation Program for Quantum Science and Technology 2021ZD0302100, and the Fundamental Research Funds for the Central Universities.

\appendix{}

\section{Spin mixing by hyperfine interaction}\label{app-B}
Hyperfine interaction can mix nuclear spin states with $m_I, m_I\pm1, m_I\pm2$. To understand this, we note that in Eq.~(\ref{hyperfine01}), 
\begin{eqnarray} 
\hat{\mathbf{I}}\cdot \hat{\mathbf{J}} &=& \hat{I}_x \hat{J}_x +\hat{I}_y \hat{J}_y +\hat{I}_z \hat{J}_z\nonumber\\&=& \frac{1}{2} [(\hat{I}_x+i  \hat{I}_y)  (\hat{J}_x-i  \hat{J}_y)+ (\hat{I}_x-i  \hat{I}_y)  (\hat{J}_x+i  \hat{J}_y)] \nonumber\\&& +\hat{I}_z \hat{J}_z\nonumber\\&\equiv& \frac{1}{2} (\hat{I}_+ \hat{J}_-+ \hat{I}_- \hat{J}_+ )  +\hat{I}_z \hat{J}_z,
\end{eqnarray}
where 
\begin{eqnarray} 
&&\hat{I}_+ \hat{J}_-|m_J, m_I\rangle = \sqrt{(I-m_I)(I+m_I+1)}\nonumber\\
&&~~\cdot\sqrt{(J+m_J)(J-m_J+1)} |m_J-1, m_I+1\rangle\nonumber\\&&~~\equiv a(m_Jm_I) |m_J-1, m_I+1\rangle,\nonumber\\
&&\hat{I}_- \hat{J}_+|m_J, m_I\rangle= \sqrt{(I+m_I)(I-m_I+1)}\nonumber\\
&&~~\cdot\sqrt{(J-m_J)(J+m_J+1)} |m_J+1, m_I-1\rangle\nonumber\\&&~~\equiv b(m_Jm_I) |m_J+1, m_I-1\rangle,\nonumber\\
&&\hat{I}_z \hat{J}_z|m_J, m_I\rangle = m_Jm_I|m_J, m_I\rangle.\nonumber
\end{eqnarray}
In the three equations above, $m_J\geq -J+1, m_I\leq I-1$ in the first equation, and $m_J\leq J-1, m_I\geq -I+1$ in the second equation. The term $(\hat{\mathbf{I}}\cdot \hat{\mathbf{J}})^2$ in Eq.~(\ref{hyperfine01}) can be expanded as
\begin{eqnarray} 
(\hat{\mathbf{I}}\cdot \hat{\mathbf{J}})^2 &=& \hat{I}_z^2 \hat{J}_z^2\nonumber\\&& +\frac{1}{2}  [(\hat{I}_+ \hat{J}_-+ \hat{I}_- \hat{J}_+ ) \hat{I}_z \hat{J}_z + \hat{I}_z \hat{J}_z(\hat{I}_+ \hat{J}_-+ \hat{I}_- \hat{J}_+ ) ]
\nonumber\\&& + \frac{1}{4}  (\hat{I}_+^2 \hat{J}_-^2+ \hat{I}_-^2 \hat{J}_+^2 + \hat{I}_+ \hat{I}_-\hat{J}_- \hat{J}_+ + \hat{I}_- \hat{I}_+\hat{J}_+ \hat{J}_-)  ,\nonumber\\
\end{eqnarray}
where 
\begin{eqnarray} 
 \hat{I}_+ \hat{I}_-\hat{J}_- \hat{J}_+ &=& (I+m_I)(I-m_I+1)\nonumber\\
 &&\cdot(J-m_J)(J+m_J+1)
\end{eqnarray}
when $m_J<J$ and $m_I>-I$, and 
\begin{eqnarray}
 \hat{I}_- \hat{I}_+\hat{J}_+ \hat{J}_- &=& (I-m_I)(I+m_I+1)\nonumber\\
 &&\cdot(J+m_J)(J-m_J+1) 
\end{eqnarray}
when $m_J>-J$ and $m_I<I$.

\section{Rabi frequencies}\label{app-A}
In this appendix, we list the dipole matrix elements found in literature for the atomic transitions involved in the model studied here, and discuss the achievable Rabi frequencies for the transitions used in the cooling of the nuclear spin qubits. 

The spontaneous emission from $(5s5p)^1P_1$ to $(5s^2)^1S_0$ has a decay rate $\Gamma=2.0\times10^8~s^{-1}$~\cite{Xu2003,Millen2010}, which is related with the dipole-transition matrix element in the context of the Weisskopf-Wigner approximation~[see Eq.~(11.33) of Ref.~\cite{DASteck}]
\begin{eqnarray}
 \Gamma_{\text{p}} &=& \frac{\omega_0^3}{9\pi \epsilon_0\hbar c^3} |\langle  [5s^2] ^1S_0 || \mathbf{d}||  [5s5p]^1P_1\rangle|^2,\label{Weisskopf}
\end{eqnarray}
 where $\mathbf{d}$ is the atomic dipole operator, $\epsilon_0$ is the free-space dielectric permittivity, $c$ is the light speed in vacuum, $\hbar$ is the Planck constant, and $\omega_0/2\pi\approx6.51\times10^{14}$~Hz is the transition frequency, which lead to $ |\langle [5s5p]^1P_1 || \mathbf{d}|| [5s^2] ^1S_0\rangle|=5.38ea_0$. This estimate should have overestimated the value of $ \langle [5s5p]^1P_1 || \mathbf{d}|| [5s^2] ^1S_0\rangle$ because the decay rate of $(5s5p)^1P_1$ is not only from the coupling between it and the ground state, but also from the coupling between it and $(5s4d)^1D_2$. Indeed, a value of about $5.25ea_0$ was suggested in Refs.~\cite{Porsev2008,Cooper2018}. The above analyses show that Eq.~(11.33) of Ref.~\cite{DASteck} is useful for the estimation of dipole matrix elements.

The transition from $(5s5p)^1P_1$ to $(5s6s)^1S_0$ is with a wavelength of 1124.232~nm~\cite{Rubbmark1978}. As in Eq.~(\ref{Weisskopf}), we have 
\begin{eqnarray}
 \Gamma_{\text{s}} &=& \frac{\omega_0^3}{\pi \epsilon_0\hbar c^3} |\langle [5s5p]^1P_1 || \mathbf{d}|| [5s6s]^1S_0 \rangle|^2.\label{Weisskopf2}
\end{eqnarray}
With $\Gamma_{\text{s}}=1.86\times10^7$s$^{-1}$~\cite{PhysRevA.46.1248}, we estimate
$|\langle [5s5p]^1P_1 || \mathbf{d}|| [5s6s]^1S_0 \rangle|=2.09ea_0$. With this value and $\pi$-polarized laser for the transition, a Rabi frequency $\Omega_{\text{ps}}=\mathcal{E}|\langle  [5s6s]^1S_0 || \mathbf{d}|| [5s5p]^1P_1 \rangle|=2\pi\times300$~MHz would require an electric field $\mathcal{E}= 1.12\times10^4V/m$, which corresponds to a beam intensity $16.7$~W$/$cm$^2$, or a laser power of $0.21$~mW if the radius of the laser spot at the atom is $20~\mu$m. This estimate shows that in principle a GHz-scale $\Omega_{\text{ps}}$ is realizable with a laser power over 2~mW.    

We did not find data about the transition probability from $(5s15d)^1D_2$ to $(5s5p)^1P_1$ in literature. One can estimate the dipole matrix element by using Coulomb wave functions as done in Refs.~\cite{Walker2008,Covey2019pra}. By the angular momentum coupling rules one can find $\langle [5s5p]^1P_1 || \mathbf{d}|| [5s15d]^1D_2 \rangle = \langle 5p || \mathbf{d}|| 15d \rangle$, where~\cite{Walker2008}
\begin{eqnarray}
|\langle 5p || \mathbf{d}|| 15d \rangle| \approx \sqrt{2}\int r P_{\text{5p}}(r)P_{\text{15d}}(r)dr,\label{radialINT}
\end{eqnarray}
which is about $0.092ea_0$ by using the effective principal quantum numbers for the 5p and 15d states suggested in Ref.~\cite{PhysRevA.52.4416}. With this estimate, A Rabi frequency $2\pi\times144.27$~MHz would require an electric field $\mathcal{E}= 1.23\times10^5V/m$, which corresponds to a laser power of $25.1$~mW if the radius of the laser spot at the atom is $20~\mu$m~(this should be experimentally feasible for the UV laser, for a power about 30~mW with a 316.6~nm UV laser was achieved for exciting Rydberg states of $^{88}$Sr in Ref.~\cite{Madjarov2020}). Note that the method via Eq.~(\ref{radialINT}) can be not as accurate in the two-electron atoms as in the alkali-metal atoms such as rubidium or cesium. Let us examine if this estimate is acceptable when we would like to argue that the Rabi frequency for the 424~nm laser field can be around $2\pi\times140$~MHz as in this paper. Note that the highest $(5snd)^1D_2$ state with transition probability to $(5s5p)^1P_1$ studied is with $n=9$~\cite{PhysRevA.46.1248}. The dipole matrix element $|\langle 5p || \mathbf{d}|| nd \rangle|$ extracted by using Eq.~(11.33) of Ref.~\cite{DASteck} via the data from Ref.~\cite{PhysRevA.46.1248} is $0.21ea_0$ for $n=9$, while the method as in Eq.~(\ref{radialINT}) leads to $0.094ea_0$ for $n=9$ by using the effective principal quantum numbers suggested in Ref.~\cite{PhysRevA.52.4416}. This means that the estimate by Eq.~(\ref{radialINT}) is likely to be smaller than the actual value, which further means that the above estimate about the required value of $\Omega_{\text{pd}}$ is within experimental feasibility.

\section{Hamiltonian matrix for numerical simulation}\label{app-C}
The theory depends on different AC Stark shifts for different $m_J$ states. To numerically investigate them, we detail the Hamiltonian for the states. The state $|[5s5p~^1P_1]-1, \uparrow\rangle$ is optically excited from the state $|[5s5p~^3P_0]0, \uparrow\rangle$ via an intermediate state with an effective Rabi frequency $\Omega_{\text{eff}}$, but is further coupled by hyperfine interaction to $|[5s5p~^1P_1]0, \downarrow\rangle$. The state $|[5s5p~^1P_1]-1, \downarrow\rangle$ is optically excited from the state $|[5s5p~^3P_0]0, \downarrow\rangle$, and is not coupled with other $|[5s5p~^1P_1], m_J, m_I\rangle$ states because $|[5s5p~^1P_1]-1, \downarrow\rangle$ has the maximal $m_J+m_I$. To suppress the hyperfine induced spin mixing, namely, the coupling between $|[5s5p~^1P_1]-1, \uparrow\rangle$ and $|[5s5p~^1P_1]0, \downarrow\rangle$, a strong $\pi$ polarized laser field is used to couple $|[5s5p~^1P_1]0, \downarrow\rangle$ and $|[5s6s~^1S_0]0, \downarrow\rangle$ with a Rabi frequency $\Omega_{\text{ps}}$~(for brevity, we assume all laser Rabi frequencies real in this paper). There is a differential energy shift between $[5s5p~^1P_1]-1, \downarrow\rangle$ and $|[5s5p~^1P_1]-1, \uparrow\rangle$. To effectively remove it so as to remove frequency resolution in the spontaneous emission, a highly detuned laser field can couple the $5s5p~^1P_1$ state with $|[5s15d~^1D_2]F, m_F\rangle$ via a 424.2399~nm~\cite{Rubbmark1978} laser. The hyperfine interaction constants are $(A,Q)/2\pi=(-194,-75)$~MHz for $5s15d~^1D_2$ as determined experimentally~\cite{PhysRevA.47.4725}, from which we find that the energies~($/\hbar$) are about $2\pi\times(-1765,-463, 604,1453,2100)$ for $F=(13/2, 11/2,9/2, 7/2, 5/2)$. When we use left-hand polarized laser field to couple $[5s5p~^1P_1]0, \downarrow\rangle, |[5s5p~^1P_1]-1, \downarrow\rangle,|[5s5p~^1P_1]-1, \uparrow\rangle$ with $5s15d~^1D_2$, only the states $[5s15d~^1D_2]F, m_F\rangle$ with $F=13/2, 11/2$ are coupled. We label the Rabi frequency for coupling $|[5s15d~^1D_2]F=13/2, m_F=-13/2\rangle$ and $|[5s5p~^1P_1]-1, \downarrow\rangle$ by $\Omega_{\text{pd}}$, then the Rabi frequencies for $\{|[5s15d~^1D_2]F=13/2, m_F=-11/2\rangle\leftrightarrow|[5s5p~^1P_1]0, \downarrow\rangle, |[5s15d~^1D_2]F=13/2, m_F=-11/2\rangle\leftrightarrow|[5s5p~^1P_1]-1, \uparrow\rangle, |[5s15d~^1D_2]F=11/2, m_F=-11/2\rangle\leftrightarrow|[5s5p~^1P_1]0, \downarrow\rangle, |[5s15d~^1D_2]F=11/2, m_F=-11/2\rangle\leftrightarrow|[5s5p~^1P_1]-1, \uparrow\rangle\}$ are $\{\xi_0,\xi_1,\xi_2,\xi_3\}\Omega_{\text{pd}}$, where $\xi_j$, $j=0-3$ are angular momentum factors. With a $\sigma^-$-polarized laser, we have 
\begin{widetext}
\begin{eqnarray} 
\langle  [5s15d~^1D_2]F, m_F |\mathbf{d}| [5s5p~^1P_1]m_J, m_I\rangle &\propto& \sum_{m_J'} C_{m_J (-1)m_J' }^{11J'}C_{m_J'm_I m_F}^{J'IF}\label{app-CG01}
\end{eqnarray}
where $J'=2$ and $m_J'\in\{-J', -J'+1,\cdots, J'\}$ are the total electron angular momentum and its $\mathbf{z}$-projection of the $5s15d~^1D_2$ state, from which we find
\begin{eqnarray} 
 \{\xi_0,\xi_1,\xi_2,\xi_3\} &=& \{\sqrt{2/13},~3/\sqrt{13},~3/\sqrt{26},~-2/\sqrt{13} \} .\label{defineXi}
\end{eqnarray}

The state $5s6s~^1S_0$ can be populated via the excitation of $|[5s5p~^1P_1]0, \downarrow\rangle$ in the cooling scheme, and $5s6s~^1S_0$ decays to the state $5s5p~^1P_1$ at a rate $18.6\times10^6$~s$^{-1}$~\cite{PhysRevA.46.1248}, which means that when excited from $|[5s5p~^1P_1]0, \downarrow\rangle$, the state $|[5s6s~^1S_0]0, \downarrow\rangle$ can decay to $|[5s5p~^1P_1]0, \downarrow\rangle$,~$|[5s5p~^1P_1]1, \downarrow\rangle$, or $|[5s5p~^1P_1]-1, \downarrow\rangle$ via emission of $\pi$, $\sigma^-$, or $\sigma^+$ polarized photons. However, due to that there is a large AC Stark shift for $|[5s5p~^1P_1]0, \downarrow\rangle$, it is barely populated, leading to negligible population in $5s6s~^1S_0$. As a result, the population in $|[5s5p~^1P_1]1, \downarrow\rangle$ is negligible. For this reason, we do not consider the laser excitation of $|[5s5p~^1P_1]1, \downarrow\rangle$ when we analyze the AC Stark shift for suppressing the hyperfine interaction. However, we include this state for it is involved in the decay of the $5s6s~^1S_0$ state.

In the basis of 
\begin{eqnarray} 
&&\{|[5s15d~^1D_2]F=13/2, m_F=-13/2\rangle,|[5s15d~^1D_2]F=13/2, m_F=-11/2\rangle,~|[5s15d~^1D_2]F=11/2, m_F=-11/2\rangle,\nonumber\\
&&|[5s6s~^1S_0]0, \downarrow\rangle,~|[5s5p~^1P_1]0, \downarrow\rangle,~
|[5s5p~^1P_1]-1, \uparrow\rangle~
|[5s5p~^1P_1]-1, \downarrow\rangle,|[5s5p~^3P_0]0, \uparrow\rangle,~|[5s5p~^3P_0]0, \downarrow\rangle,
\nonumber\\
&&
~|[5s^2~^1S_0]0, \uparrow\rangle,~
|[5s^2~^1S_0]0, \downarrow\rangle,~
|\mathscr{A}\rangle,~|[5s5p~^1P_1]1, \downarrow \rangle\},\label{basis}
\end{eqnarray}
the Hamiltonian consists of the atom-laser interaction $\hat{H}_{\text{a-l}}$,
\begin{eqnarray} 
\hat{H}_{\text{a-l}} &=& \frac{1}{2}\left(
\begin{array}{ccccccccccccc}
 2(\Delta_{\text{pd}}+\Delta) & 0& 0&  0&0&0& \Omega_{\text{pd}}& 0& 0&0& 0& 0& 0 \\ 
 0& 2(\Delta_{\text{pd}}+\Delta) &   0&0&\xi_0 \Omega_{\text{pd}}&\xi_1 \Omega_{\text{pd}}& 0&0& 0&0& 0& 0& 0 \\ 
 0&0 & 2(\Delta_{\text{pd}}+E_{\text{hf}}+\Delta) & 0& \xi_2 \Omega_{\text{pd}}&\xi_3\Omega_{\text{pd}}&0& 0& 0&0& 0& 0 & 0\\ 
 0 &  0 &0& 2\Delta& \Omega_{\text{ps}}& 0&0& 0& 0&0& 0& 0& 0 \\ 
 0&\xi_0\Omega_{\text{pd}} & \xi_2\Omega_{\text{pd}}& \Omega_{\text{ps}}& 2\Delta& 0&0& 0& 0&0& 0& 0 & 0\\ 
  0 &\xi_1 \Omega_{\text{pd}} & \xi_3 \Omega_{\text{pd}}& 0& 0& 2\Delta&0& \Omega_{\text{eff}}& 0&0& 0& 0& 0 \\ 
  \Omega_{\text{pd}} &0 & 0& 0& 0& 0&2\Delta& 0& \Omega_{\text{eff}}&0& 0& 0 & 0\\ 
 0& 0  & 0& 0& 0& \Omega_{\text{eff}}&0& 0& 0&0& 0& 0 & 0\\ 
 0& 0  & 0& 0& 0& 0&\Omega_{\text{eff}}& 0& 0&0& 0& 0 & 0\\ 
 0 & 0& 0& 0& 0&0& 0& 0&0& 0& 0 & 0& 0 \\ 
 0 & 0& 0& 0& 0&0& 0& 0&0& 0& 0 & 0& 0 \\ 
 0 & 0& 0& 0& 0&0& 0& 0&0& 0& 0 & 0 & 0 \\ 
 0 & 0& 0& 0& 0&0& 0& 0&0& 0& 0 & 0 & 0
\end{array}
\right),\label{app-Hmatrix}
\end{eqnarray} 
the Zeeman shift $\hat{H}_{\text{Zee}}=g_J\mu_{\text{B}} J_z B -g_I\mu_{\text{n}} I_z B $~(we ignore the Zeeman shift for the three $5s15d~^1D_2$ states and $|\mathscr{A} \rangle$), 
and the hyperfine interaction $\hat{h} $ which couples $|[5s5p~^1P_1]-1, \uparrow\rangle$ and $|[5s5p~^1P_1]0, \downarrow\rangle$, where $E_{\text{hf}}=2\pi\times1.3$~GHz as shown above Eq.~(\ref{app-CG01}); here the detuning is defined as the dipole transition frequency deducted by the laser frequency. The matrix element of the hyperfine interaction is
\begin{eqnarray} 
&&\langle m_J', m_I' |\hat{h}  |m_J, m_I\rangle\nonumber\\
&&=\langle m_J', m_I' | A\hat{\mathbf{I}}\cdot \hat{\mathbf{J}}+ Q\frac{ 3(\hat{\mathbf{I}}\cdot \hat{\mathbf{J}})^2+ 1.5\hat{\mathbf{I}}\cdot \hat{\mathbf{J}} -IJ(I+1)(J+1) }{2IJ(2I-1)(2J-1)}  |m_J, m_I\rangle\nonumber\\
&&= \delta_{m_Jm_J'}\delta_{m_Im_I'}\bigg [A m_Im_J  +Q\frac{ 3m_I^2m_J^2+ 1.5 m_Im_J  -IJ(I+1)(J+1) }{2IJ(2I-1)(2J-1)}    \nonumber\\
&&+   Q\frac{ 3(I+m_I)(I-m_I+1)(J-m_J)(J+m_J+1)\Theta(J-m_J)\Theta(m_I+I)
}{8IJ(2I-1)(2J-1)}   \nonumber\\
&&+  Q\frac{ 3(I-m_I)(I+m_I+1)(J+m_J)(J-m_J+1) \Theta(J+m_J)\Theta(I-m_I)
}{8IJ(2I-1)(2J-1)}   \bigg]\nonumber\\
&&+\delta_{m_J'(m_J-1)}\delta_{m_I'(m_I+1)} \bigg [\frac{1}{2}A a(m_Im_J)  +Q\frac{ 1.5(m_Im_J+ (m_I+1)(m_J-1) )a(m_Im_J) + 0.75 a(m_Im_J)  }{2IJ(2I-1)(2J-1)}  \bigg]\nonumber\\
&&+ \delta_{m_J'(m_J+1)}\delta_{m_I'(m_I-1)}\bigg [\frac{1}{2}A b(m_Im_J)  +Q\frac{ 1.5(m_Im_J+ (m_I-1)(m_J+1) )b(m_Im_J) + 0.75 b(m_Im_J)  }{2IJ(2I-1)(2J-1)}  \bigg]\nonumber\\
&&+\delta_{m_J'(m_J-2)}\delta_{m_I'(m_I+2)} \bigg [Q\frac{ 3a(m_Im_J)a\big((m_I+1)(m_J-1)\big)   }{8IJ(2I-1)(2J-1)}  \bigg]\nonumber\\
&&+\delta_{m_J'(m_J+2)}\delta_{m_I'(m_I-2)} \bigg [Q\frac{ 3b(m_Im_J)b\big((m_I-1)(m_J+1)\big)   }{8IJ(2I-1)(2J-1)}  \bigg],\label{h-Hamiltonian}
\end{eqnarray}
where $\Theta(x)=1$ if $x>0$ and $\Theta(x)=0$ if $x\leq 0$, and
\begin{eqnarray} 
a(m_Jm_I) &=& \sqrt{(I-m_I)(I+m_I+1)}\sqrt{(J+m_J)(J-m_J+1)} ,\nonumber\\
b(m_Jm_I) &=& \sqrt{(I+m_I)(I-m_I+1)} \sqrt{(J-m_J)(J+m_J+1)}  .\nonumber
\end{eqnarray}
The hyperfine constants are $(A,Q)/2\pi=(-3.4,~39)$~MHz~\cite{Kluge1974} for $(5s5p)^1P_1$. 

\end{widetext}

%

\end{document}